\documentclass[amssymb,amsmath,superscriptaddress]{revtex4}
\usepackage{natbib,graphicx,subfigure,subeqnarray,fancyhdr,amsmath,epstopdf,multirow,color}
\pdfoutput=1 
\begin{document}

\renewcommand{\labelitemi}{$-$}

\newcommand{\Fc}{\mathcal{F}}\newcommand{\Rc}{\mathcal{R}}\newcommand{\dd}{\mathrm{d}}
\newcommand{\ee}{\mathrm{e}}\newcommand{\ci}{\mathrm{i}}\newcommand{\ib}{\mathbf{i}}
\newcommand{\jb}{\mathbf{j}}\newcommand{\kb}{\mathbf{k}}\newcommand{\ab}{\mathbf{a}}
\newcommand{\Fb}{\mathbf{F}}\newcommand{\fb}{\mathbf{f}}\newcommand{\Gb}{\mathbf{G}}
\newcommand{\Mb}{\mathbf{M}Ä}\newcommand{\nb}{\mathbf{n}}\newcommand{\Sb}{\mathbf{S}}
\newcommand{\Sbs}{\mathbf{S^*}}\newcommand{\Rb}{\mathbf{R}}\newcommand{\Sigb}{\boldsymbol{\Sigma}}
\newcommand{\Sigbs}{\boldsymbol{\Sigma^*}}
\newcommand{\omegab}{\boldsymbol{\omega}}
\newcommand{\epsb}{\boldsymbol{\epsilon}}
\newcommand{\ub}{\mathbf{u}}
\newcommand{\eb}{\mathbf{e}}\newcommand{\vv}[1]{\underline{#1}}\newcommand{\ev}{\vv{e}}
\newcommand{\rv}{\vv{r}}\newcommand{\TT}[1]{\underline{\underline{#1}}}\newcommand{\omb}{\mathbf{\omega}}
\newcommand{\Ub}{\mathbf{U}}\newcommand{\xb}{\mathbf{x}}\newcommand{\rb}{\mathbf{r}}
\newcommand{\ssb}{\mathbf{s}}\newcommand{\Xb}{\mathbf{X}}\newcommand{\Rey}{\mbox{\textit{Re}}}
\newcommand{\mean}[1]{\langle #1\rangle}
\newcommand{\ddp}{[p]^\pm}\newcommand{\taub}{\mbox{\boldmath$\tau$}}\newcommand{\Fr}{\mbox{\textit{Fr}}}
\let\grad\nabla\newcommand{\z}{\zeta}\newcommand{\kk}{\kappa}\newcommand{\tkk}{\tilde{\kappa}}
\newcommand{\e}{\varepsilon}\newcommand{\zb}{\bar{\zeta}}\let\grad\nabla\let\bcdot\cdot
\newcommand{\half}{{\textstyle\frac{1}{2}}}
\newcommand{\textfrac}[2]{{\textstyle\frac{#1}{#2}}}
\newcommand{\LF}[1]{{#1}^{\mathrm{LF}}}\newcommand{\Lap}[1]{{#1}^{\mathrm{L}}}
\newcommand{\ds}{*\!*}\newcommand{\cond}[2]{\frac{\mathrm{D} #1}{\mathrm{D} #2}}
\newcommand{\pard}[2]{\frac{\partial #1}{\partial #2}}\newcommand{\totd}[2]{\frac{\mathrm{d}#1}{\mathrm{d}#2}}
\newcommand{\pardd}[3]{\frac{\partial^2 #1}{\partial #2 \partial #3}}
\newcommand{\Real}{\mbox{Re}}\newcommand{\Imag}{\mbox{Im}}
\newcommand{\Fpint}{=\!\!\!\!\!\!\!\int}
\newcommand{\txi}{\tilde\xi}\newcommand{\dxi}{\delta\xi}
\newcommand{\tpsi}{\tilde\psi}\newcommand{\dpsi}{\delta\psi}
\newcommand{\change}[1]{\textcolor{red}{ #1}}
\makeatletter
\def\sgn{\mathop{\operator@font sgn}}
\makeatother

\title{Efficiency optimization and symmetry-breaking in a  model of ciliary locomotion}
\author{S\'ebastien Michelin}
\email{sebastien.michelin@ladhyx.polytechnique.fr}
\affiliation{Department of Mechanical and Aerospace Engineering, University of California San Diego, 9500 Gilman Drive, La Jolla CA 92093-0411.}
\affiliation{LadHyX -- D\'epartement de M\'ecanique, Ecole
  polytechnique, 91128 Palaiseau Cedex, France}
\author{Eric Lauga}
\email{elauga@ucsd.edu}
\affiliation{Department of Mechanical and Aerospace Engineering, University of California San Diego, 9500 Gilman Drive, La Jolla CA 92093-0411.}

\date{\today}
\begin{abstract}
A variety of swimming microorganisms, called ciliates, exploit the bending of a large number of small and densely-packed  organelles, termed cilia, in order to propel themselves in a viscous fluid. We consider a spherical envelope model for such ciliary locomotion where the dynamics of the individual cilia are replaced by that of a continuous overlaying surface allowed to deform tangentially to itself. Employing a variational approach, we determine numerically the time-periodic deformation of such surface which leads to low-Reynolds locomotion with  minimum rate of energy dissipation (maximum efficiency). Employing both  Lagrangian and Eulerian points of views, we show that in the optimal swimming stroke, individual cilia display weak asymmetric beating, but that a significant symmetry-breaking occurs at the organism level, with the whole surface deforming in a wave-like fashion reminiscent of metachronal waves of biological cilia. This wave motion is analyzed using a formal modal decomposition, is found to occur in the same direction as the swimming direction, and is interpreted as due to a spatial distribution of phase-differences in the kinematics of individual cilia. Using additional constrained optimizations, as well a constructed analytical ansatz, we derive a complete optimization diagram where all swimming efficiencies, swimming speeds, and amplitude of surface deformation can be reached, with the  mathematically optimal swimmer, of efficiency one half, being a  singular limit. Biologically, our work suggests therefore that metachronal waves  may allow cilia to propel cells forward while reducing  the energy dissipated in the surrounding fluid.

%Within our framework, we will thus be able to conclude that metachronal waves are hydrodynamically optimal.

\end{abstract}
\maketitle

\section{Introduction}

Swimming microorganisms are found in a large variety of environments. From  spermatozoa cells to bacteria in the human body, from unicellular protozoa to multicellular algae swimming in the ocean,  these small organisms are able to generate net locomotion by exploiting  their interaction with a surrounding viscous fluid \cite{lauga2009}. Because of their small dimensions, the Reynolds number associated with their motion, $\Rey=UL/\nu$, where $U$ is the swimming velocity, $L$ the typical size of the organism, and $\nu$ the fluid kinematic viscosity, is close to zero and the effects of  body and fluid inertia are both negligible. As a result, the motion of swimming microorganisms is based on the exploitation of viscous drag to generate thrust through periodic non-time-reversible shape changes \cite{lighthill1975, childress1981, lauga2009}. 

\begin{figure}
\begin{center}
\includegraphics[width=10cm]{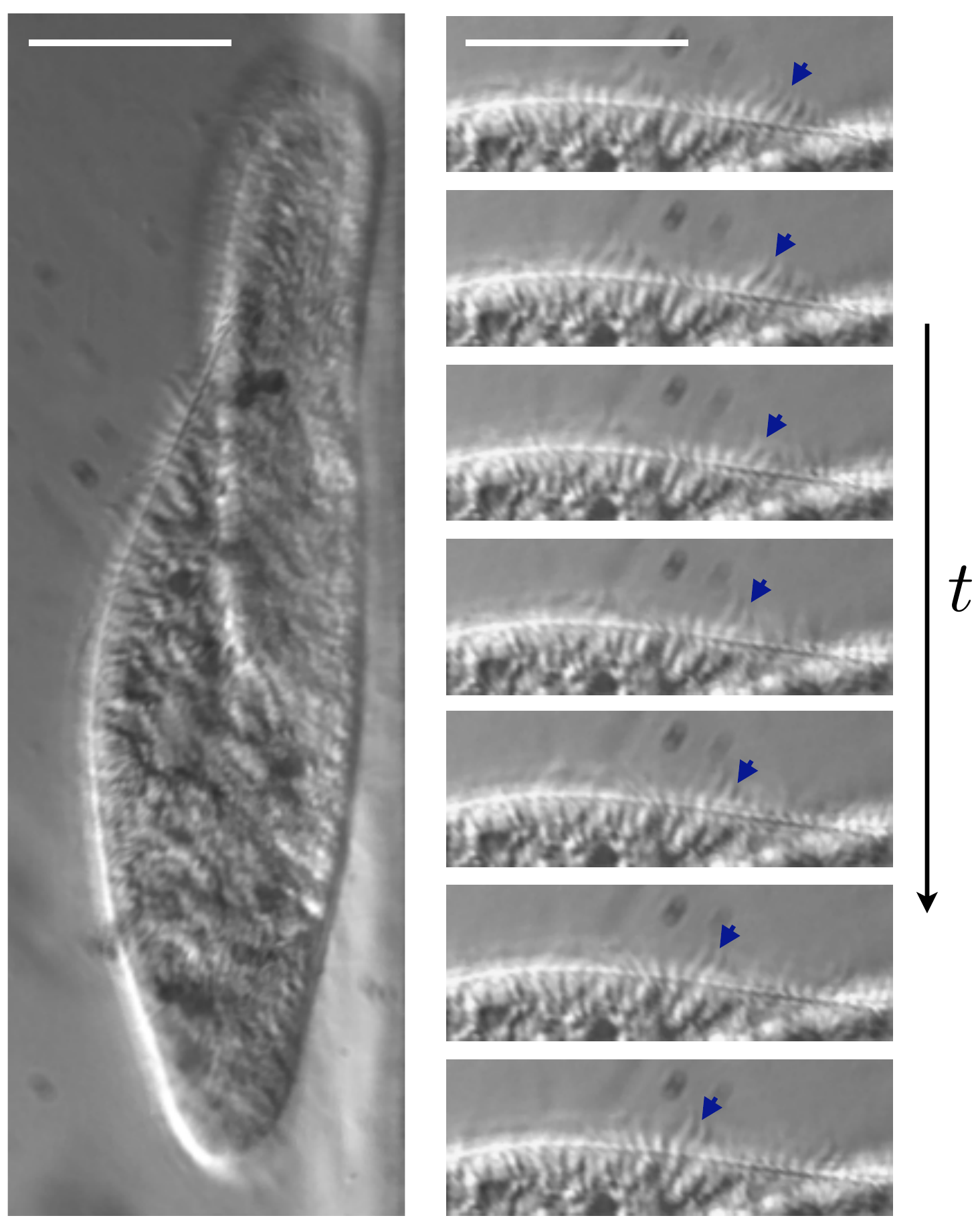}
\caption{
{\it Paramecium} swimming using metachronal waves (Hoffman modulation contrast microscopy, 40X magnification). 
Left: Picture of the cell in a microchannel. 
Right: Visualization of metachronal waves of cilia deformation propagating along the cell; time increases from top to bottom,  with a time difference of  1/300 s between each picture; arrows indicate the propagation of a wave of effective strokes in the cilia array.  
Scale bar is 50 $\mu$m in both pictures. 
Pictures courtesy of Sunghwan (Sunny) Jung, Virginia Tech.
}
\label{picture}
\end{center}
\end{figure}

Many swimming microorganisms use the beating of elongated flexible appendages attached to their surface to produce motion \cite{brennen1977}. These appendages are known as cilia or flagella depending on their distribution density on the cell, and their size relative to that of the organism. Eukaryotic flagella, such as those used by invertebrate or mammalian  spermatozoa, are longer than the cell head and are only found in small numbers (typically one for spermatozoa, and a few for other eukaryotes \cite{lauga2009}). In contrast, most of the surface of ciliates such as \emph{Paramecium} (see Fig.~\ref{picture}, left) is covered by cilia  much shorter than the cell body \cite{blake74}. Ciliary motion is also functionally essential for respiratory systems, where the beating of cilia covering lung epithelium permits the transport of mucus and foreign particles out of the respiratory tract \cite{sleigh1988}. In that case, the cilia support is fixed and the cilia motion produces a net flow. 

Eukaryotic flagella and cilia have similar diameters (of the order of 200 nm), beating frequencies (of the order of 10 Hz) and internal structure. They differ however in length. The typical length of a sperm cell flagellum  is of the order of 50 $\mu$m, while cilia, such as those  of \emph{Paramecium}, have a typical length of 5--10 $\mu$m \cite{brennen1977}. Both eukaryotic flagella and cilia are subject to distributed actuation through the sliding of neighboring polymeric filaments (microtubules doublets)  \cite{brokaw1972,brokaw1989}.  Notably, bacterial flagella, although bearing the same name, are much smaller (typically 20 nm diameter and 5---10 $\mu$m in length), have a much simpler internal structure than their eukaryotic equivalent, and are passively driven by a rotating motor located in the cell wall \cite{berg1973,berg2004}.

The propagation of wave patterns is essential to both flagellar and ciliary propulsion \cite{gibbons1981}. The deformation of individual eukaryotic flagella in planar or helical wave patterns leads to non-time-reversible kinematics, and is  thus responsible for the net swimming motion of the associated microorganism \cite{gray1955b,lighthill1975, purcell1977,childress1981, lauga2009}. In this paper, we focus on ciliary propulsion, for which two levels of symmetry-breaking (and two types of waves) are observed \cite{brennen1977}. At the level of an individual cilium, a deformation wave propagates along the cilium length as for flagellar motion. The beating pattern is however different from that of individual flagella, and the individual stroke of a cilium  can be decomposed into two parts:  an effective stroke, during which the cilium is extended and offers the most resistance to the fluid, and a recovery stroke, in which the cilium is bent in such a way as to reduce the viscous drag. 

In addition to such asymmetric beating at the level of an individual cilium, the beating coordination of neighboring cilia at the organism level  results in a collective behavior known as metachronal waves.  All cilia on the surface of a microorganism perform similar beating patterns, but they deform in time with a small phase difference with respect to their neighbors, and these phase differences are spatially-distributed in a way that leads to symmetry-breaking at the level of the whole cell, and the formation of a wave pattern of surface deformation \cite{blake1974b,brennen1977} (see Fig.~\ref{picture}, right)
. The origin of the synchronization responsible for the metachronal waves in ciliary propulsion is still debated, but several recent studies have suggested that it results from hydrodynamic interactions between neighboring cilia \cite{gueron1997,vilfan2006,lenz2006,guirao2007,niedermayer2008}. 

In this paper, we  consider the energy cost and hydrodynamic efficiency associated with ciliary propulsion. By efficiency, we understand here a relative measure of the organism displacement or velocity to the energy dissipated through viscous stresses in the flow to produce this motion. Of course,  this differs from the actual energetic cost for the organism which includes  metabolism and other internal biological considerations, and we limit ourselves here to the purely hydrodynamical aspect of the efficiency. Although little is known experimentally about the actual energy consumption associated with ciliary propulsion, some studies suggest that metachronal waves  reduce the energy loss \cite{gueron1999}. More generally, the question of the hydrodynamic efficiency of different swimming patterns is at the heart of many low-$\Rey$ locomotion investigations of the extent to which the swimming modes observed in nature could be optimal with respect to hydrodynamic efficiency. In particular,  numerous studies have considered the optimal beating of flagella \cite{pironneau74,lighthill1975, tam07,tam2008thesis,spagnolie2010}. Through a theoretical and numerical  optimization framework, we determine in this paper the particular collective cilia beating patterns that minimize the dissipation of mechanical energy.

Computing the flow around a ciliated organism accurately is difficult because of the large number of appendages  deforming and  interacting. Two types of modeling approaches have been proposed in the past to address this problem. The first type of model, termed sublayer modeling, considers the dynamics of individual cilia, either theoretically in a simplified fashion   \cite{blake74, brennen1977}, or numerically with all hydrodynamic interactions \cite{gueron1992,gueron1993}. The study of the collective dynamics of a large number of cilia remains however a costly computation \cite{gueron1997,gueron1999}.

An alternative approach, motivated by the densely packed arrangement of the cilia on the surface of the organism, is based on the description of the swimmer by a deformable, continuous surface enveloping the cilia at each instant \cite{taylor1951,blake1971,brennen1977}. This so-called envelope model substitutes the motion of material surface points (cilia tips) with the deformation of the continuous surface, and is expected to be a good approximation when the density of cilia is sufficiently high. 
In the particular case of a purely spherical swimmer, this model is known as a squirmer \cite{lighthill1952,blake1971}, and it was recently used to study the collective dynamics and the rheology of suspensions of model swimming microorganisms \cite{ishikawa2006,ishikawa2007a}. Although some organisms using cilia do have a spherical shape (e.g. \emph{Volvox} \cite{drescher2009}), most ciliated microorganisms have an elongated body. Nonetheless, the squirmer model and the spherical approximation allow one to reduce the complexity of the problem in order to shed some light on the  fundamental properties of symmetry-breaking in ciliary locomotion.

In this paper, we thus consider locomotion by a squirmer in a Newtonian fluid without inertia as a model for locomotion of a spherical ciliated cell. We will assume that the squirmer can deform its shape tangentially in a time-periodic fashion, and hence the shape remains that of a sphere for all times. For a given time-periodic stroke --- that is for a given periodic Lagrangian surface displacement field --- we are able to compute the swimming velocity of the swimmer, the energy dissipation in the fluid, and use both to define the stroke hydrodynamic efficiency. 
We will not restrict our analysis to small-amplitude deformations \cite{shapere1989b,stone1996}, but will allow arbitrary large-amplitude tangential deformations to take place. The purpose of our work is to then determine the optimal squirming stroke which maximizes this swimming efficiency (or, alternatively, minimizes the amount of work done against the fluid for a fixed average swimming velocity), and to study its characteristics. In a first part, we will consider the general problem of determining the optimal stroke theoretically and numerically. In a second part, we will consider the limitation to the surface kinematics  introduced by the finite size of the cilia, and how it affects the optimal swimming stroke. In all cases, we will show that the optimal swimming strokes show very little asymmetry at the level of individual cilia (Lagrangian framework) but display strong symmetry-breaking at the level of the whole organism (Eulerian framework), reminiscent of phase differences between cilia and of metachronal waves observed in biology. Within our framework, we will thus be able to conclude that metachronal waves are hydrodynamically optimal.

In Sec.~\ref{sec:unconstr}, after a brief review of the squirmer model and its dynamics, the optimization procedure is presented, togetherw with the resulting optimal stroke. In Sec.~\ref{sec:constr}, a constraint is added to the optimization to take into account the finite-length of the cilia and limit the surface displacement. Based on the observations of the optimal strokes obtained in Secs.~\ref{sec:unconstr} and \ref{sec:constr}, an analytical ansatz is constructed in Sec.~\ref{sec:ansatz} that achieves asymptotically the theoretical upper bound for the swimming efficiency of a squirmer. In Sec.~\ref{sec:results}, the physical properties of the optimal strokes are presented and discussed, in particular the wave characteristics. Our results are finally summarized and discussed in  Sec.~\ref{sec:conclusions}.

%%%%%%%%%%%%%%%%%%%%%%%%%%%%
% NEW SECTION: UNCONSTRAINED OPTIMIZATION
%%%%%%%%%%%%%%%%%%%%%%%%%%%%

\section{Optimal swimming stroke of spherical swimmer}
\label{sec:unconstr}

We consider in this paper the dynamics of a spherical microorganism able to produce locomotion by imposing time-periodic tangential displacements of its spherical surface --- the  so-called squirmer approximation  --- as an envelope model for ciliated cells.  Only  axisymmetric strokes with no azimuthal displacements are considered, thereby restricting the swimming motion to a pure time-varying translation.

\subsection{Equations of motion and swimming efficiency}

\begin{figure}
\begin{center}
\includegraphics[width=10cm]{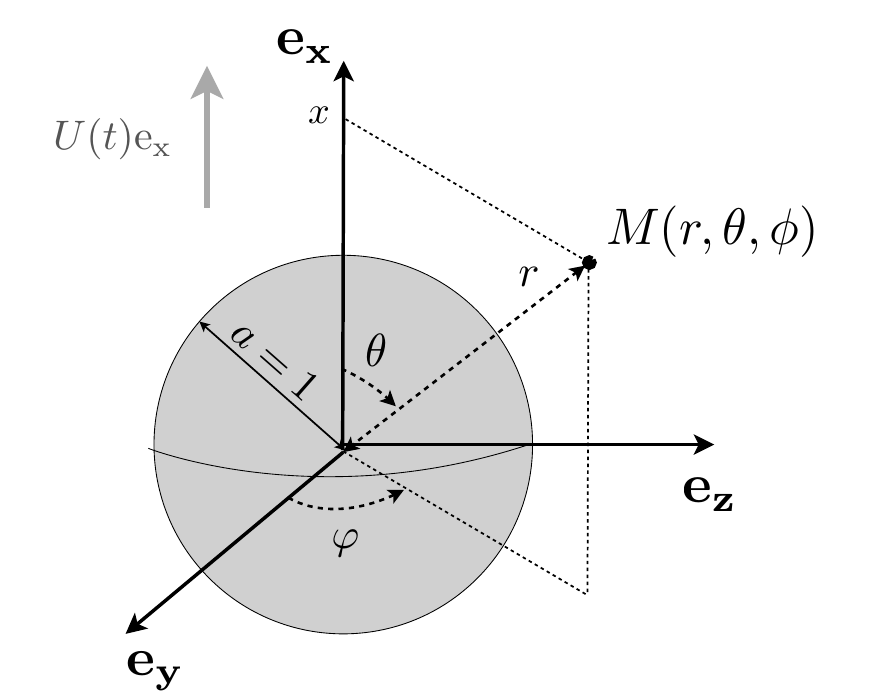}
\caption{Notation for the squirmer model. A co-moving frame centered on the squirmer center is used with spherical polar coordinates $(r,\theta,\varphi)$. The axisymmetry of the surface motion imposes a purely translating swimming motion along $\eb_x$ with time-varying velocity $U(t)$. See text for details. }\label{fig:notations}
\end{center}
\end{figure}

\subsubsection{Geometric description}
The dynamics of the micro-swimmer is studied in a translating reference frame centered at the squirmer  center. In this co-moving frame, the fluid and surface motions are described in spherical polar coordinates $(r,\theta,\varphi)$ (see notation in Fig.~\ref{fig:notations}). All quantities are non-dimensionalized using the swimmer radius $a$, the stroke frequency $f$ and the fluid dynamic viscosity $\mu$. 

The surface of the squirmer ($r=1$) is described by two Lagrangian variables $0\leq\theta_0\leq\pi$ and $0\leq\varphi_0\leq 2\pi$, and the position of each surface element is defined by the time evolution of its polar and azimuthal angles $\theta(\theta_0,\varphi_0,t)$ and $\varphi(\theta_0,\varphi_0,t)$. Considering purely axisymmetric deformations of the swimmer's surface, we have
\begin{equation}\label{eq:geomax}
\theta=\vartheta(\theta_0,t),\quad \varphi=\varphi_0.
\end{equation}
In the following, we will therefore omit the dependence in $\varphi_0$, all particles located on a circle of constant $\theta$ having the same behavior. By a straightforward symmetry argument, the swimming velocity of the organism must thus be of the form $U(t)\eb_x$.

Each point on the swimmer surface can be equivalently described by its polar angle $\theta$ or its vertical cartesian coordinate $x=\cos\theta$. In the following, both notations will be used to describe the surface motion 
\begin{equation}
\theta=\vartheta(\theta_0,t),\,\, \mbox{or}\,\,\, x=\xi(x_0,t).
\end{equation}
Most of the analytical work will be done in terms of the $x$ variable as equations adopt a simpler form in this case. Physical properties of the stroke will however be discussed using the variable $\theta$ which describes the actual angular displacement of the cilia tips on the surface.

The Lagrangian label $x_0=\cos\theta_0$ is chosen such that $x_0=-1$ ($\theta_0=\pi)$ and $x_0=1$ ($\theta_0=0$) are respectively the south and north poles of the squirmer (Fig.~\ref{fig:notations}). Spatial boundary conditions at both poles, as well as time-periodicity impose that
\begin{align}\label{eq:bcpoles}
\xi(-1,t)&=-1\quad \textrm{and}\quad \xi(1,t)=1 \quad \textrm{for}\quad 0\leq t\leq 2\pi,\\
\xi(x_0,t)&=\xi(x_0,t+2\pi)\quad \textrm{for}\quad -1\leq x_0\leq 1.\label{eq:period}
\end{align}

Note that there exist multiple choices for the Lagrangian label $x_0$ (or $\theta_0$), and it is not necessarily the position of the material point at $t=0$. Any other Lagrangian label $x_1$ satisfying Eq.~\eqref{eq:bcpoles} can be used provided that $x_0\rightarrow x_1$ is a bijective function. In the following, we use the mean value of $\vartheta$ over a swimming period as Lagrangian label and $\theta_0=\mean{\vartheta}$, where $\mean{\cdot}$ denotes the averaging operator in time.

Finally, in a discrete representation of the three-dimensional  motion of cilia, it would possible for cilia to cross. This is however forbidden in the continuous axisymmetric envelope model used here, and the surface velocity is uniquely defined at each point; $\xi(x_0)$ must therefore be a bijective function, which leads to
\begin{equation}\label{eq:bijection}
\pard{\xi}{x_0}> 0,\quad\textrm{for    } -1\leq x_0\leq 1,\,\,\,0\leq t\leq 2\pi.
\end{equation}

In the co-moving frame, the surface velocity $\ub^S$ is purely tangential $\ub^S=u_\theta^S(\theta,t)\eb_\theta$, and is related to the surface displacement by a partial time derivative
\begin{equation}\label{eq:motioneq}
\pard{\vartheta}{t}(\theta_0,t)=u^S_\theta(\vartheta(\theta_0,t),t).
\end{equation}
Equivalently, this equation can be rewritten using the variables $(x,t)$ as
\begin{equation}\label{eq:motioneq_x}
\pard{\xi}{t}(x_0,t)=u(\xi(x_0,t),t),\qquad\textrm{with   }u(x,t)=-\sqrt{1-x^2}u_\theta^S(\cos^{-1}x,t).
\end{equation}
The tangential velocity $u_\theta^S(\theta,t)$ is the velocity {at a fixed point} $\theta$, while $\partial \vartheta/\partial t$ is the tangential velocity {of a given material point} labelled by $\theta_0$. Equation \eqref{eq:motioneq} is therefore the fundamental conversion from Eulerian to Lagrangian quantities. Note that $u(x,t)$ in Eq.~\eqref{eq:motioneq_x} is the axial component of the surface velocity (along $\eb_x$).

\subsubsection{Swimming motion of a squirmer}
In the limit of zero Reynolds number $\Rey=0$, the equations for the incompressible flow around the squirmer simplify into the non-dimensional Stokes equations
\begin{equation}\label{eq:stokes}
\grad^2\ub=\grad p,\quad \nabla.\ub=0,
\end{equation}
with boundary conditions, expressed in the co-moving frame, as
\begin{align}\label{eq:bcfluid_1}
\ub&=\ub^S=u_\theta^S(\theta,t)\eb_\theta\quad\textrm{for   }r=1,\\
\ub&=-U\eb_x\qquad \textrm{for    }r\rightarrow\infty.\label{eq:bcfluid_2}
\end{align}
Eqs.~\eqref{eq:stokes}--\eqref{eq:bcfluid_2} can be solved explicitely  as \citep{blake1971}
\begin{align}
u_r&=-U(t)\cos(\theta)+\alpha_1(t)\frac{L_1(\cos\theta)}{r^3}+\frac{1}{2}\sum_{n=2}^\infty\left[\frac{1}{r^{n+2}}-\frac{1}{r^n}\right](2n+1)\alpha_n(t)L_n(\cos\theta)\label{eq:ur},\\
u_\theta&=U\sin\theta+\frac{\alpha_1(t)}{2}\frac{V_1(\theta)}{r^3}+\frac{1}{4}\sum_{n=2}^\infty\left[\frac{n}{r^{n+2}}-\frac{n-2}{r^n}\right](2n+1)\alpha_n(t)V_n(\theta)\label{eq:uth},\\
p&=-\sum_{n=2}^\infty\left(\frac{4n^2-1}{n+1}\right)\alpha_n(t)\frac{L_n(\cos\theta)}{r^{n+1}}\label{eq:p},
\end{align}
where $L_n(x)$ is the $n$-th Legendre polynomial, $V_n(\theta)$ is defined as
 \begin{equation}\label{eq:Vndef}
V_n(\theta)=-\frac{(2n+1)\sin\theta}{n(n+1)}L_n'(\cos\theta),
\end{equation}
and the time coefficients $\alpha_n(t)$ are obtained uniquely from the expansion of the surface tangential velocity in spherical harmonics
\begin{equation}\label{eq:alphndef}
{u_\theta(\theta,t)=u^S_\theta(\theta,t)=}\sum_{n=1}^\infty\alpha_n(t)V_n(\theta).
\end{equation}

In the Stokes regime, the inertia of the swimmer is negligible so the fluid force is zero at all times
\begin{equation}\label{eq:swim}
\Fb_\textrm{ext}=\int_S\left[-p\,\nb+(\grad\ub+\grad\ub^T)\cdot\nb\right]\dd S=0.
\end{equation}
The viscous torque is also zero by symmetry.
From Eqs. \eqref{eq:ur}--\eqref{eq:p} and \eqref{eq:swim}, the swimming velocity $U$ as well as the instantaneous rate of work of the swimmer on the fluid can then be obtained as \citep{blake1971}
\begin{align}\label{eq:swimvel}
U(t)&=\alpha_1(t),\\
\label{eq:power}
\mathcal{P}(t)&=12\pi\left[\alpha_1(t)^2+\frac{1}{3}\sum_{n=2}^\infty\frac{(2n+1)^2}{n(n+1)}\alpha_n(t)^2\right].
\end{align}
The swimming velocity is completely determined by the first mode $\alpha_1(t)$, which is thereafter referred to as the swimming mode. All others modes, $\alpha_n(t)$, do not contribute to the swimming velocity but do contribute to the energy consumption, and in that sense are penalizing the swimming efficiency of the organism. However, the existence of these non-swimming modes is required to ensure the periodicity of the surface displacement.

\subsubsection{Swimming efficiency}

In this paper we are going to derive the optimal stroke kinematics, and thus we have to define our cost function. Since low-$\Rey$ locomotion is essentially  a geometrical  problem 
\cite{purcell1977,shapere1987}, the appropriate cost function is basically a way to normalize this geometrical problem. 
Here we will use the traditional definition of a low-$\Rey$ swimming efficiency, $\eta$, given by 
\begin{equation}\label{eq:eff_def}
\eta=\frac{\mean{U}\mathcal{T}^*}{\mean{\mathcal{P}}},
\end{equation}
where $\mathcal{T}^*$ is the force required to drag a rigid body of same shape as the swimmer at the time-averaged swimming velocity $\mean{U}$, and $\mean{\mathcal{P}}$ the mean rate of energy dissipation in the fluid during swimming \citep{lighthill1975, childress1981, tam2008thesis,chattopadhyay06, lauga2009}. 
The cost function, $\eta$, given in Eq.~\eqref{eq:eff_def}, has traditionally been termed an efficiency, but perhaps more accurately it should be termed a normalization, as it is the ratio between the average rate of work necessary to move the body in two different ways: in the numerator, dragging the body with an external force, and in the denominator, self-propelled motion at the same speed. It is thus not a thermodynamic efficiency, and for general swimmers does not have to be less then one, although for most biological cells it is on the order of $1\%$ \cite{lauga2009}.

In the particular case of a squirmer, Stokes' formula gives $\mathcal{T}^*=6\pi\mean{U}$, and the efficiency (Eq.~\ref{eq:eff_def}) can be obtained analytically as
\begin{equation}\label{eq:efficiency}
\eta=\frac{6\pi\mean{U}^2}{\mean{\mathcal{P}}}=\frac{\mean{\alpha_1}^2}{2\left[\mean{\alpha_1(t)^2}+\frac{1}{3}\displaystyle\sum_{n=2}^\infty\frac{(2n+1)^2}{n(n+1)}\mean{\alpha_n(t)^2}\right]}\cdot
\end{equation}
From the Cauchy--Schwartz inequality, we see that Eq.~\eqref{eq:efficiency} leads to an upper bound of $\eta\leq 1/2$. This upper bound, which we will  discuss in more detail in Sec. \ref{sec:ansatz}, is tighter than the one obtained in Ref.~\cite{stone1996} ($\eta\leq 3/4$) and corresponds to the treadmilling microswimmer described in  Ref.~\cite{leshansky2007}. Importantly, the definition of efficiency retained here is purely mechanical and does not characterize the absolute efficiency of the locomotion mode for the organism, which is influenced by many other factors, including feeding and metabolic costs. In addition, since we are using an envelope model for the cilia, our approach does not allow us to capture the fluid dissipation in the sublayer (i.e. near the cilia).

Using Eqs. \eqref{eq:motioneq_x} and \eqref{eq:alphndef}, we have the dynamics
\begin{equation}
\pard{\xi}{t}=-\sqrt{1-\xi^2}\sum_{m=1}^\infty \alpha_m(t)V_m(\cos^{-1}\xi)=-(1-\xi^2)\sum_{m=1}^\infty\frac{(2m+1)\alpha_m(t)L_m'(\xi)}{m(m+1)}\cdot
\end{equation}
Multiplying both sides of the previous equation by $L_n'(\xi)$ and integrating in $\xi$ leads to
\begin{equation}\label{eq:alphn_comp}
\alpha_n(t)=-\frac{1}{2}\int_{-1}^1L_n'(\xi)\pard{\xi}{t}\dd \xi=-\frac{1}{2}\int_{-1}^1L_n'(\xi(x_0,t))\pard{\xi}{t}\pard{\xi}{x_0}\dd x_0.
\end{equation}
Integrating by part and using the boundary condition at the poles (Eqs. \ref{eq:bcpoles}) finally leads to the Lagrangian-Eulerian relationship
\begin{equation}\label{eq:alphan}
\alpha_n(t)=\frac{1}{2}\int_{-1}^1L_n(\xi(x_0,t))\pard{^2\xi}{x_0\partial t}\dd x_0.
\end{equation}

For a given stroke $x=\xi(x_0,t)$, the coefficients $\alpha_n(t)$ are computed from Eq.~\eqref{eq:alphan} and used to determine the swimming efficiency $\eta[\xi]$ using Eq.~\eqref{eq:efficiency}.

\subsection{Optimization of the swimming efficiency}
The objective of the present work is to determine the  stroke $\xi(x_0,t)$ maximizing the swimming efficiency $\eta[\xi]$. Physically, the optimal stroke will therefore be the one swimming the {furthest for a given amount of dissipated energy}, or, alternatively,  will be the one minimizing the rate of work done by the swimmer for a given swimming speed. To characterize the optimal stroke, we use  a variational approach.

\subsubsection{Swimming efficiency gradient in the stroke functional space}
Let us consider a particular reference swimming stroke $\txi(x_0,t)$, and the perturbed stroke $\txi(x_0,t)+\dxi(x_0,t)$, where $\dxi(x_0,t)$ is a small perturbation to the reference stroke. In the following, tilde quantities correspond to the reference (or, initial) stroke. From the boundary conditions (Eq.~\ref{eq:bcpoles}) applied to $\txi$ and $\txi+\dxi$, we obtain the following boundary conditions and periodicity constraints on $\dxi(x_0,t)$
\begin{align}
\label{eq:bc_dxi}
\dxi(-1,t)=\dxi(1,t)=0 &\textrm{   for all  } 0\leq t\leq 2\pi,\\
\dxi(x_0,t)=\dxi(x_0,t+2\pi) &\textrm{  for all  }-1\leq x_0\leq 1\label{eq:per_dxi}.
\end{align}
Retaining only the linear contribution, the change in swimming efficiency is given by
\begin{align}\label{eq:delta_eta}
\delta\eta=\eta[\txi+\dxi]-\eta[\txi]&=\frac{\mean{\tilde\alpha_1}\delta\mean{\alpha_1}}{D}-\frac{\mean{\tilde\alpha_1}^2\left(\delta\mean{\alpha_1^2}+\displaystyle\frac{1}{3}\sum\limits_{n=2}^\infty\frac{(2n+1)^2}{n(n+1)}\delta\mean{\alpha_n^2}\right)}{2D^2},
\end{align}
with
\begin{equation}
 D=\mean{\tilde\alpha_1^2}+\frac{1}{3}\sum\limits_{n=2}^\infty\frac{(2n+1)^2}{n(n+1)}\mean{\tilde\alpha_n^2}.\end{equation}
From Eqs. \eqref{eq:alphan} and \eqref{eq:bc_dxi}, the perturbation $\delta\alpha_n(t)$ induced on $\alpha_n(t)$ is computed as 
\begin{equation}\label{eq:delta_alphan}
\delta\alpha_n(t)=\frac{1}{2}\int_{-1}^1\left(L_n'(\txi)\pardd{\txi}{x_0}{t}\dxi-L_n'(\txi)\pard{\txi}{x_0}\pard{\dxi}{t}\right)\dd x_0.
\end{equation}
For $n=1$ and using Eq.~\eqref{eq:per_dxi}, the time average of Eq.~\eqref{eq:delta_alphan} leads to 
\begin{equation}\label{eq:delta_0alpha1}
\delta\mean{\alpha_1}=\frac{1}{2\pi}\int_0^{2\pi}\int_{-1}^1\dxi\cdot\pardd{\txi}{x_0}{t}\dd x_0\dd t.
\end{equation}
Keeping only the leading order contribution, we have
\begin{equation}
\delta\mean{\alpha_n^2}=2\mean{\tilde\alpha_n\,\delta\alpha_n}.
\end{equation}
Using Eq.~\eqref{eq:delta_alphan} and integration by part in time, we find
\begin{equation}\label{eq:delta_0ean_alphasquare}
\delta\mean{\alpha_n^2}=\frac{1}{2\pi}\int_0^{2\pi}\int_{-1}^1\dxi\cdot\left[\tilde\alpha_n(t)\left(2L'_n(\txi)\pardd{\txi}{x_0}{t}+L_n''(\txi)\pard{\txi}{x_0}\pard{\txi}{t}\right)+\dot{\tilde\alpha}_nL_n'(\txi)\pard{\txi}{x_0}\right]\dd x_0\dd t.
\end{equation}
After substitution into Eq.~\eqref{eq:delta_eta}, the variation $\delta\eta$ can finally  be written as
\begin{equation}\label{eq:delta_eta_2}
\delta\eta=\int_{-1}^1\int_0^{2\pi}F[\txi](x,t)\dxi(x_0,t)\dd x_0\dd t, 
\end{equation}
where $F[\txi](x_0,t)$ is the efficiency gradient in the stroke functional space evaluated at $\txi$
\begin{align}\label{eq:grad_xi}
F[\txi]&=\frac{\mean{\tilde\alpha_1}}{4\pi D}\left\{2\pardd{\txi}{x_0}{t}-\frac{\mean{\tilde\alpha_1}}{D}\left[\left(2\tilde\alpha_1\pardd{\txi}{x_0}{t}+\dot{\tilde\alpha}_1\pard{\txi}{x_0}\right)\right.\right.\\
&+\left.\left.\frac{1}{3}\sum_{n=2}^\infty\frac{(2n+1)^2}{n(n+1)}\left(2L'_n(\txi)\left[\tilde\alpha_n\pardd{\txi}{x_0}{t}+\dot{\tilde\alpha}_nL_n'(\txi)\pard{\txi}{x_0}\right]+L_n''(\txi)\pard{\txi}{x_0}\pard{\txi}{t}\right)\right]\right\}\nonumber\cdot
\end{align}

\subsubsection{Projection on the subspace of the boundary conditions}
$F[\txi]$ is the gradient of $\eta$ with respect to the swimming stroke $\txi$. Choosing a sufficiently small $\dxi$ aligned with $F[\txi]$ guarantees an increase of the swimming efficiency ($\delta\eta>0$) between $\txi$ and $\txi+\dxi$ as the integrand in Eq.~\eqref{eq:delta_eta_2} is positive everywhere, which suggests the manner in which we can numerically iterate to find the optimal swimmer.   However, such a change $\dxi$ along $F[\txi]$ does not guarantee that $\txi+\dxi$ will satisfy the boundary and periodicity conditions (Eqs. \ref{eq:bcpoles}--\ref{eq:period}) or the constraint on monotoneous variations (Eq.~\ref{eq:bijection}). To circumvent this difficulty, we notice that for any stroke $\xi(x_0,t)$, there exists at least one continuous function $\psi(x_0,t)$ such that we can write
\begin{equation}\label{eq:psidef}
\xi(x_0,t)=-1+\displaystyle\frac{2\int_{-1}^{x_0}\left[\psi(x',t)\right]^2\dd x'}{
\int_{-1}^1\left[\psi(x',t)\right]^2\dd x'}\cdot
\end{equation}
Performing the optimization on the field $\psi(x_0,t)$ rather than $\xi(x_0,t)$ frees us from imposing the boundary, periodicity and bijection conditions separately, as, through
Eq.~\eqref{eq:psidef}, any $\psi(x_0,t)$ will generate an acceptable $\xi(x_0,t)$. The analysis of the previous paragraph remains valid and we now must evaluate the variation $\dxi(x_0,t)$ induced by a small change $\dpsi(x_0,t)$. From Eq.~\eqref{eq:psidef}, we have
\begin{equation}\label{deltaq}
\dxi(x_0,t)=4\left[\frac{\int_{-1}^{x_0}\tpsi\,\dpsi\,\dd x'}{\int_{-1}^1\tpsi^2\,\dd x'}-\frac{\left(\int_{-1}^1\tpsi\,\dpsi\,\dd x'\right)\left(\int_{-1}^{x_0}\tpsi^2\,\dd x'\right)}{\left(\int_{-1}^1\tpsi^2\,\dd x'\right)^2}\right]\cdot
\end{equation}
Substitution of this result into Eq.~\eqref{eq:delta_eta_2} leads, after integration by part and rearrangement of the result, to
\begin{equation}\label{eq:delta_eta_3}
\delta\eta=\int_0^{2\pi}\int_{-1}^1G[\tpsi](x,t)\dpsi(x_0,t)\dd x_0 \dd t,
\end{equation}
with $G[\tpsi](x_0,t)$ the efficiency gradient in the functional space of $\psi$, that can be written as:
\begin{align}\label{eq:grad_psi}
G[\tpsi](x_0,t)=\frac{4\tpsi(x_0,t)}{\left(\int_{-1}^1\tpsi^2(x',t)\dd x'\right)^2}&\left[\left(\int_{x_0}^1F[\txi](x',t)\dd x'\right)\left(\int_{-1}^1\tpsi^2(x',t)\dd x'\right)\right.\\
&-\left.\int_{-1}^1F[\txi](x',t)\left(\int_0^{x'}\tpsi^2(x'',t)\dd x''\right)\dd x'\right]\nonumber
\end{align}
with $F[\txi]$ given in Eq.~\eqref{eq:grad_xi}.

\subsection{Numerical optimization}
The optimization on the swimming stroke is performed iteratively and numerically, using a steepest ascent algorithm \cite{pozrikidis1998}. Given a guess $\psi_n(x_0,t)$, the efficiency $\eta[\psi_n]$ and the gradient $G[\psi_n]$ are computed numerically using Eqs.~\eqref{eq:efficiency}, \eqref{eq:alphan}, \eqref{eq:grad_xi} and \eqref{eq:grad_psi}. The next guess $\psi_{n+1}(x_0,t)$ is obtained numerically by marching in the functional space in the gradient direction as
\begin{equation}\label{eq:steepest}
\psi_{n+1}(x_0,t)=\psi_n(x_0,t)+\varepsilon G[\psi_n](x_0,t),
\end{equation}
where $\varepsilon$ is a small number. For small enough $\varepsilon$, Eq.~\eqref{eq:delta_eta_3} guarantees that $\delta\eta=\eta[\psi_{n+1}]-\eta[\psi_n]>0$. Starting from an initial guess $\psi_0$, and using this approach iteratively, one travels in the $(\psi,\eta)$-space along the steepest slope in $\eta$ until convergence is reached. As the purpose of the present paper is to obtain physical insight on the maximum efficiency strokes, we did not attempt to construct the fastest converging algorithm, and it is possible that faster convergence would be achieved, for example, using a variation of the conjugate gradient algorithm.

The efficiency $\eta$ and gradient $G$ are computed numerically from $\psi$ using spectral methods in both $t$ (Fast Fourier Transform) and $x_0$ (Chebyshev spectral methods). To avoid aliasing phenomena introduced by the successive non-linear products, the 2/3-dealiasing rule is applied in both space and time before taking each physical product.

Convergence is achieved when marching in the direction of the gradient $G$ does not induce any increase of the swimming efficiency anymore, even after successive reductions of the step size $\varepsilon$. This algorithm, as most iterative optimization techniques, can not guarantee the finding of an absolute maximum of the swimming efficiency but only of local maxima. Because of the infinite number of dimensions of the functional space, it is expected and observed that some optimization runs will lead to local maxima with small values of $\eta$. This difficulty is overcome by performing several runs with different initial conditions and/or resolution until convergence to similar solutions provide enough confidence in the finding of an absolute maximum.

As pointed out before, $\xi(x_0,t)$ is not a unique description of the swimming stroke as any bijection on the lagrangian label will lead to another equivalent representation of same $\eta$. To identify whether two strokes are equivalent, we must therefore resort to the comparison of their physical characteristics or velocity distribution.

\begin{figure}
\begin{center}
\includegraphics[width=10cm]{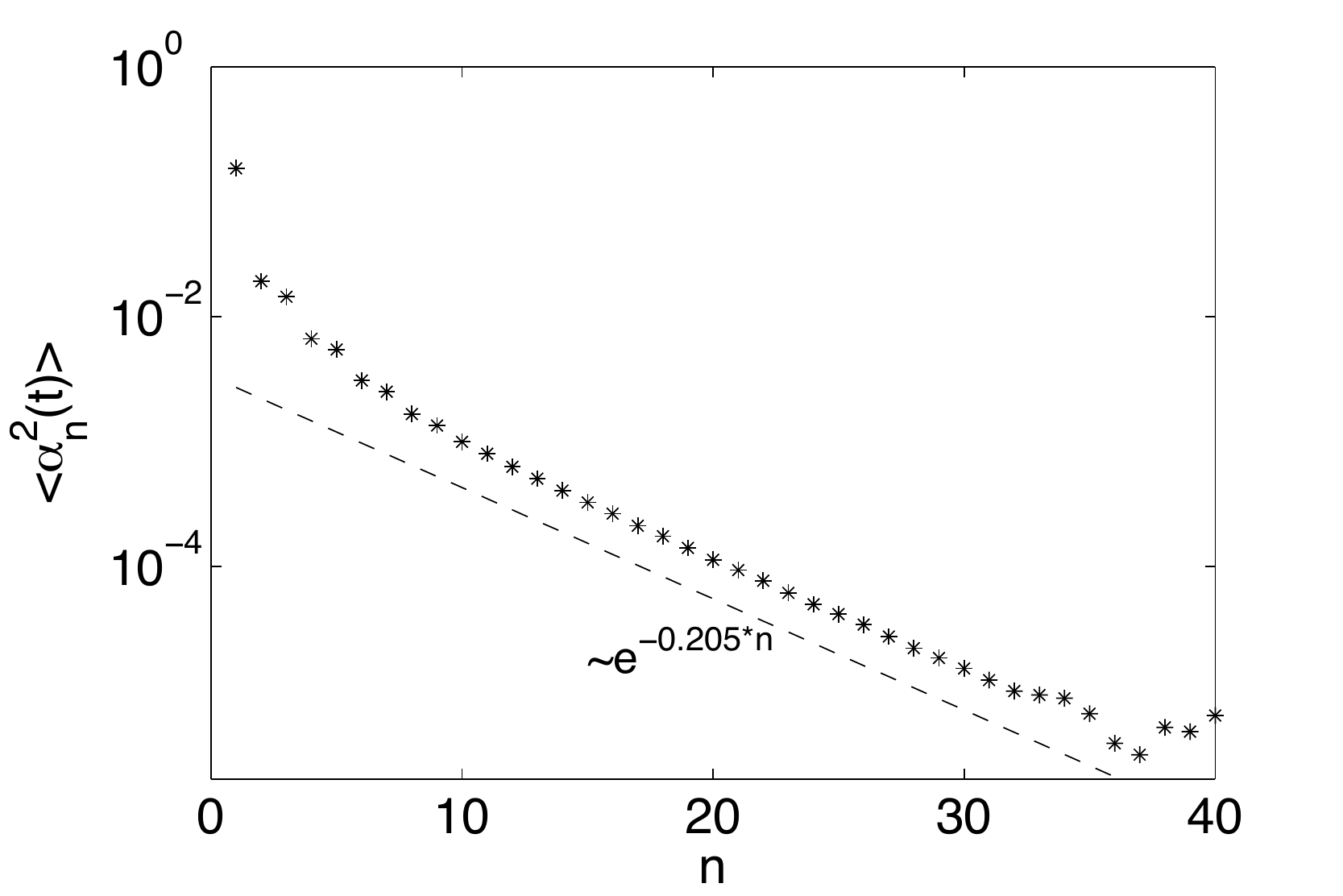}
\caption{The contribution of the successive modes, $\mean{\alpha_n^2}$, to the efficiency (stars) decreases exponentially with the mode order, $n$. The results are plotted for the unconstrained optimization with efficiency $\eta\approx 22.2\%$, and were obtained with $M=40$, $N_x=150$ and $N_t=128$.}\label{fig:malpha2}
\end{center}
\end{figure}

We use $N_x$ Chebyshev Gauss-Lobatto points  in $x_0$, and $N_t$ equidistant points in $t$. The infinite sum in Eq.~\eqref{eq:efficiency} must be truncated at $M$ modes. A large enough number of modes $M$ must be computed for the efficiency estimate to be accurate, thereby imposing a practical minimum on $N_x$ and $N_t$ for the integrals involved in Eq.~\eqref{eq:alphan} to be computed accurately. In the representative cases presented in the rest of this work, typical values of the discretization parameters were $M\approx 40$--$80$, $N_x\approx 120$--$200$ and $N_t\approx 64$--$128$. Figure \ref{fig:malpha2} shows that the contribution $\mean{\alpha_n^2}$ of mode $n$ to the efficiency of the reference stroke decreases exponentially with $n$. One can then estimate the truncation error on $\eta$ as $\approx 0.03\%$.

\subsection{Initial guesses for the swimming stroke}
Different sets of initial conditions were used to ensure convergence of the procedure. Examples of such initial conditions are
\begin{equation}\label{eq:init_cond}
\xi(x_0,t)=x_0+c_1(1-x_0^2)^N\cos(t-c_2x_0),
\end{equation}
with $N$ an integer, $c_1\ll 1$ and $0.5\leq c_2/\pi\leq 10$. Eq.~\eqref{eq:init_cond} corresponds to a traveling wave in $x_0$ of wavespeed $1/c_2$, attenuated at both poles.  The efficiency being a quadratic function of the swimming velocity (see Eq.~\ref{eq:efficiency}), the initial condition must have non-zero swimming velocity and must therefore not be time-reversible (Purcell's theorem \cite{purcell1977}). In the following, traveling wave patterns will be identified and it is important to ensure that the wave-like characteristics of the optimal solution result from the efficiency optimization and not from the particular initial guess considered. In that regard, different wave amplitudes and wave velocities were tested in Eq.~\eqref{eq:init_cond}. Also, a superposition of multiple waves traveling in different directions were tested as well as ``hemispheric solutions" defined independently in each hemisphere (the equator then becoming a fixed point), and ``elliptical solutions" of the form
\begin{equation}\label{eq:init_cond_ell}
\xi(x_0,t)=-1+\frac{(1+x_0)\left[1+c_1\sin\left(\frac{\pi(1+x_0)}{2}\right)\right]}{\sqrt{\sin^2\left(t-c_2x_0\right)+\left[1+c_1\sin\left(\frac{\pi(1+x_0)}{2}\right)\right]^2\cos^2(t-c_2 x_0)}}\cdot
\end{equation}
Independently of the particular choice of initial conditions, we obtain a systematic convergence to the results presented in the following section.

\subsection{Unconstrained results: optimal swimming stroke}
\label{uresults}
\begin{figure}
\begin{center}
\includegraphics[width=10cm]{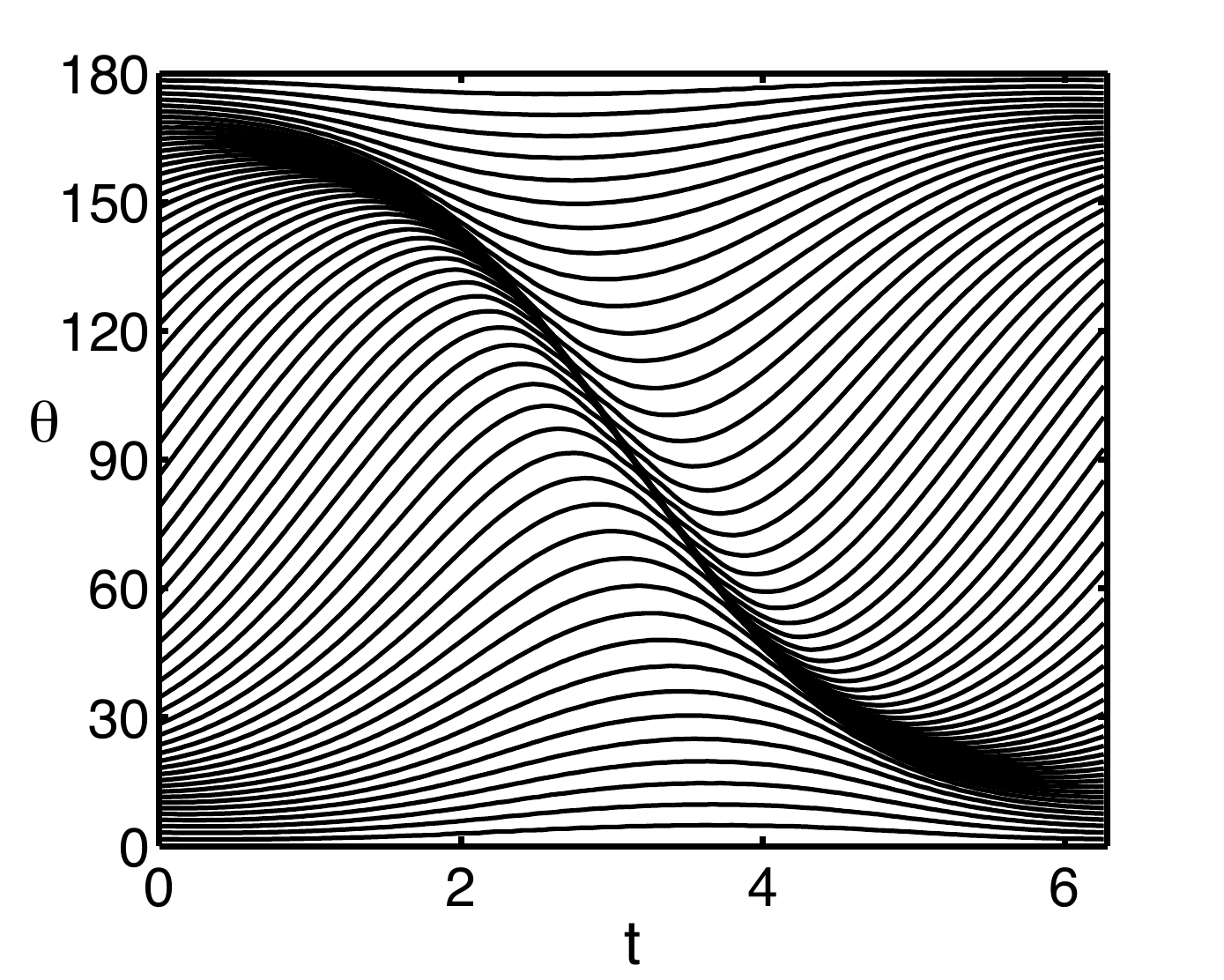}
\caption{Lagrangian description of the optimal swimming stroke with no
  constraint on the maximum displacement. The maximum amplitude of displacement is $\Theta_\textrm{max}\approx52.6^\circ$, {the mean swimming velocity is $\mean{U}\approx0.33$ } and the swimming efficiency is $\eta\approx22.2\%$. Each curve illustrates the
  trajectory $\theta=\vartheta(\theta_0,t)$ of a single material surface point.}\label{fig:theta_plot_reference}
\end{center}
\end{figure}

\begin{figure}
\begin{center}
\includegraphics[width=15cm]{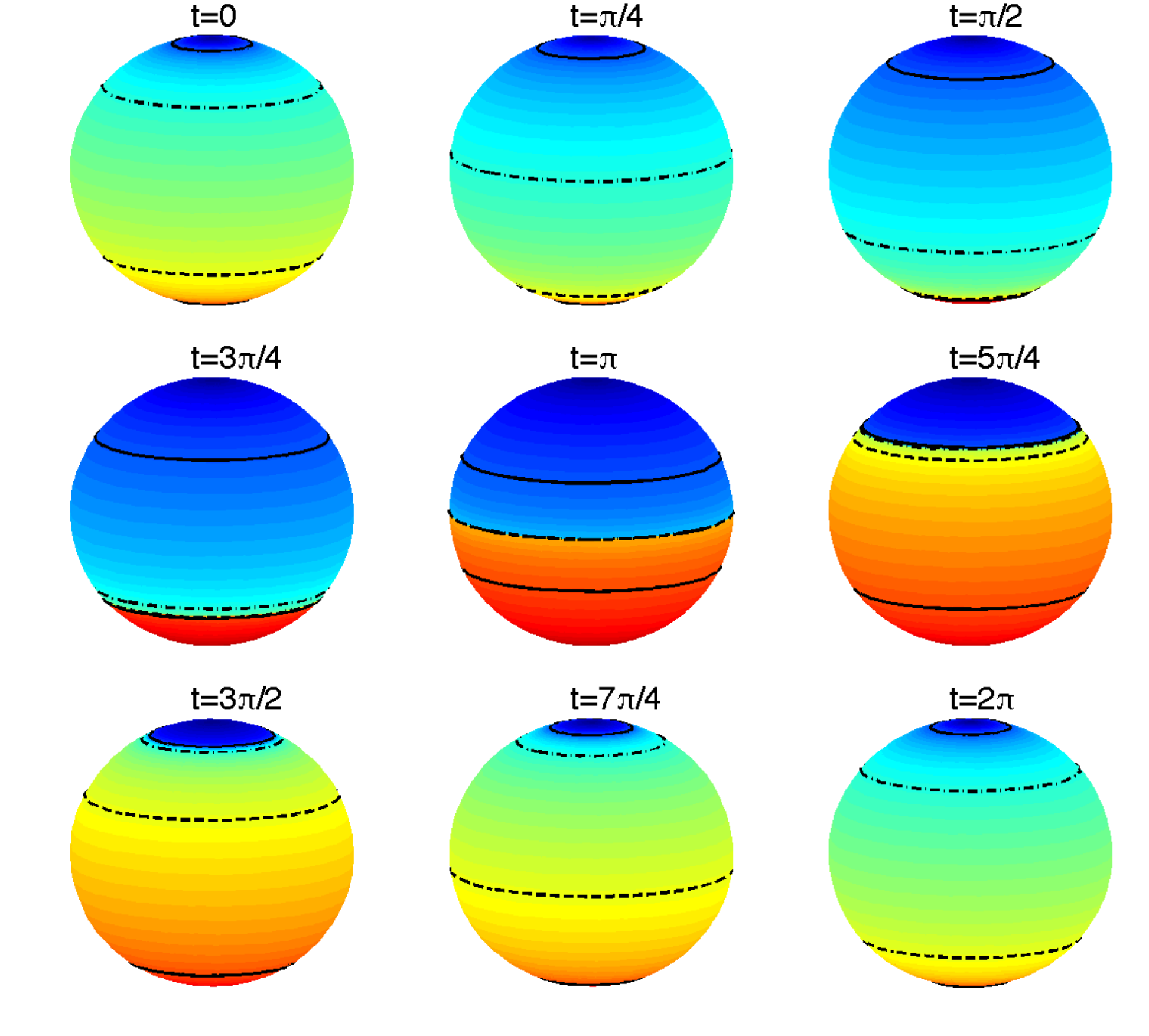}
\caption{(color online) Snapshots of the displacement of the squirmer surface over
  a swimming period in the optimal unconstrained stroke ($\Theta_\textrm{max}\approx 52.6^\circ$, {$\mean{U}\approx0.33$,} $\eta\approx22.2\%$). The surface color of each material point on
  the surface refers to the mean polar angle of that material
  point. Black lines also correspond to the location of particular material points, and have been added for clarity.}\label{fig:movie_reference}
\end{center}
\end{figure}

For all these different initial conditions, we observe the convergence of our numerical approach toward a stroke of efficiency $\eta\approx 22.2\%$.  We will refer to this solution as the optimal unconstrained stroke thereafter.

\begin{figure}
\begin{center}
\includegraphics[width=12cm]{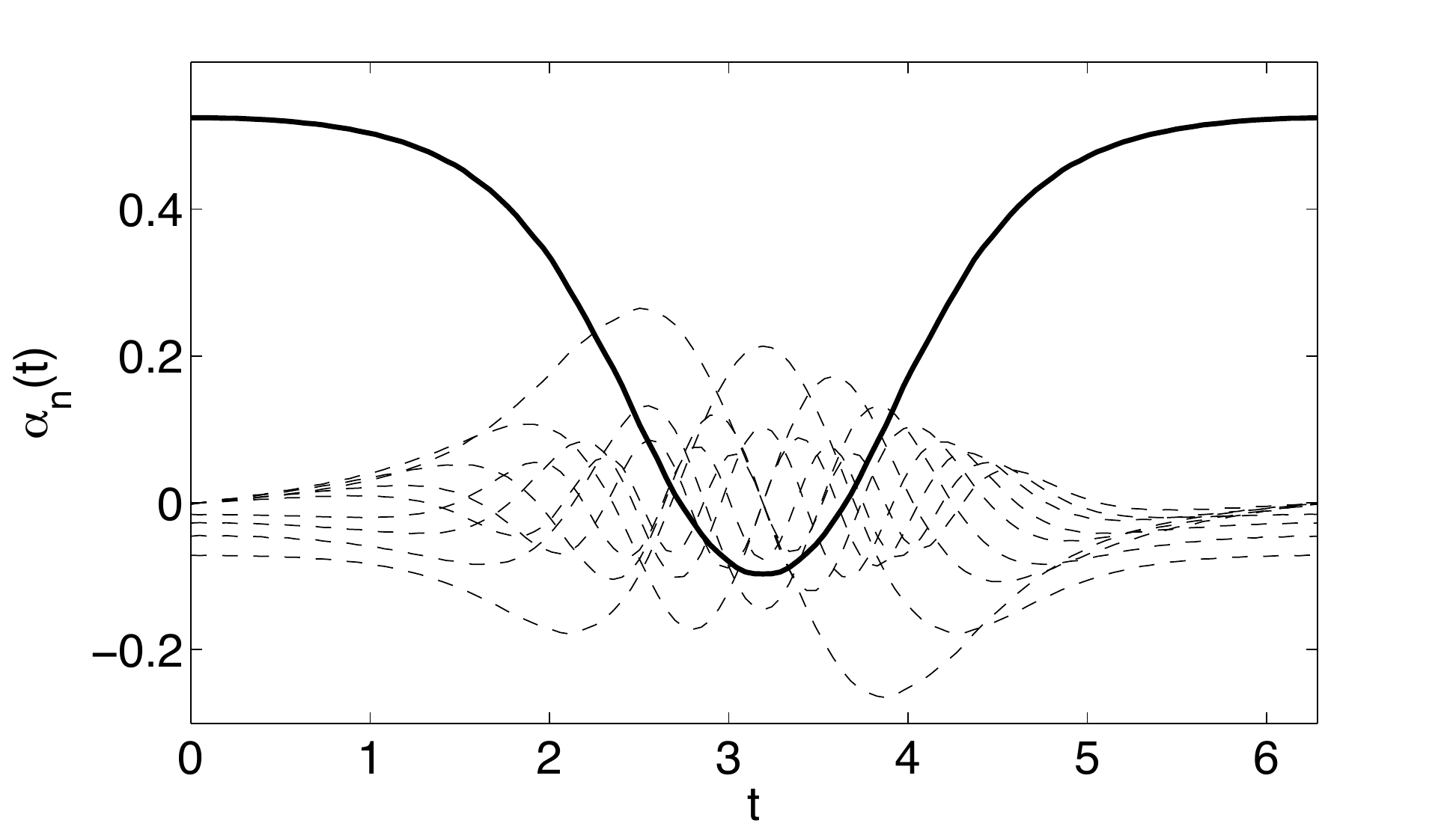}
\caption{Time-evolution of the ten first modes, $\alpha_n(t)$ ($n\leq 10$), over the period for the optimal unconstrained  stroke shown in Fig.~\ref{fig:movie_reference}. {The swimming mode or swimming velocity  $U(t)=\alpha_1(t)$ is plotted as a thick solid line}. Modes 2 through 10 are plotted as dashed line. The swimming mode is observed to dominate during the effective stroke (for $t\leq\pi/2$ and $t\geq3\pi/2$), while many modes are significant during the recovery stroke ($\pi/2\leq t\leq 3\pi/2$). }\label{fig:alpha}
\end{center}
\end{figure}

The trajectories of surface elements along the sphere $\theta=\vartheta(\theta_0,t)$ are shown on Fig.~\ref{fig:theta_plot_reference}, and Fig.~\ref{fig:movie_reference} presents a sequence of snapshots corresponding to a stroke period. On Fig.~\ref{fig:movie_reference}, the color code is a Lagrangian label and allows one to track the position of a particular surface particle in time. A few lines $\theta_0=\textrm{constant}$ are also represented for better visualization of the surface motions. The average swimming velocity associated with this stroke is directed upward and equal to $\mean{U}=0.33$. Due to the length and time scaling chosen here, this corresponds to a mean swimming velocity slightly above one body-length-per-period. {The swimming velocity is not constant throughout the period as can be seen on Fig.~\ref{fig:alpha}. }

The optimal  unconstrained swimming stroke illustrated in Figs.~\ref{fig:theta_plot_reference} and \ref{fig:movie_reference} can be decomposed, in time, into two parts: (1) an effective stroke where the surface moves downward (increasing $\theta$) while stretching; (2) 
a recovery stroke where the surface elements that migrated toward the south pole during the effective swimming stroke are brought back toward the north pole, in the same direction as the swimming velocity. The surface is highly compressed in this phase, which gives it a shock-like structure (see the dark region in  Fig.~\ref{fig:theta_plot_reference}). {The instantaneous swimming velocity, $U(t)$, is maximum and roughly constant throughout the effective stroke, while the recovery stroke is associated with a reduced (even reversed) swimming velocity (Fig.~\ref{fig:alpha}).}

These two strokes are {not exactly} successive in time as their boundary is not vertical on Fig.~\ref{fig:theta_plot_reference}.  The recovery stroke is mostly visible during the middle half of the stroke period ($\pi/2\leq t\leq 3\pi/2$) (see Figs.~\ref{fig:theta_plot_reference} and \ref{fig:movie_reference}). Outside of this time domain, the entire organism surface is moving downward (effective stroke), with the possible exception of two small regions located near the poles. This is confirmed by the predominance of the swimming mode $\alpha_1(t)$ over the other modes outside of the domain $[\pi/2\,,\,3\pi/2]$ (see Fig.~\ref{fig:alpha}).

Although individual cilia motion is not explicitly represented in this continuous  envelope model, the swimming stroke obtained through this optimization process shares many similarities with metachronal waves observed in ciliated microorganisms  \cite{brennen1977,sleigh1988}. Indeed, a small phase difference in the motion of neighboring surface points (or cilia tips) can be observed: as can be seen in Fig.~\ref{fig:theta_plot_reference}, for the range {$20^\circ\leq\theta_0\leq 160^\circ$}, the maximum and minimum of a given trajectory occur with a small time delay compared to those of a trajectory with a slightly greater value of {$\theta_0$}.

\begin{figure}
\begin{center}
\begin{tabular}{cc}
\subfigure[In-phase beating]{\includegraphics[height=6cm]{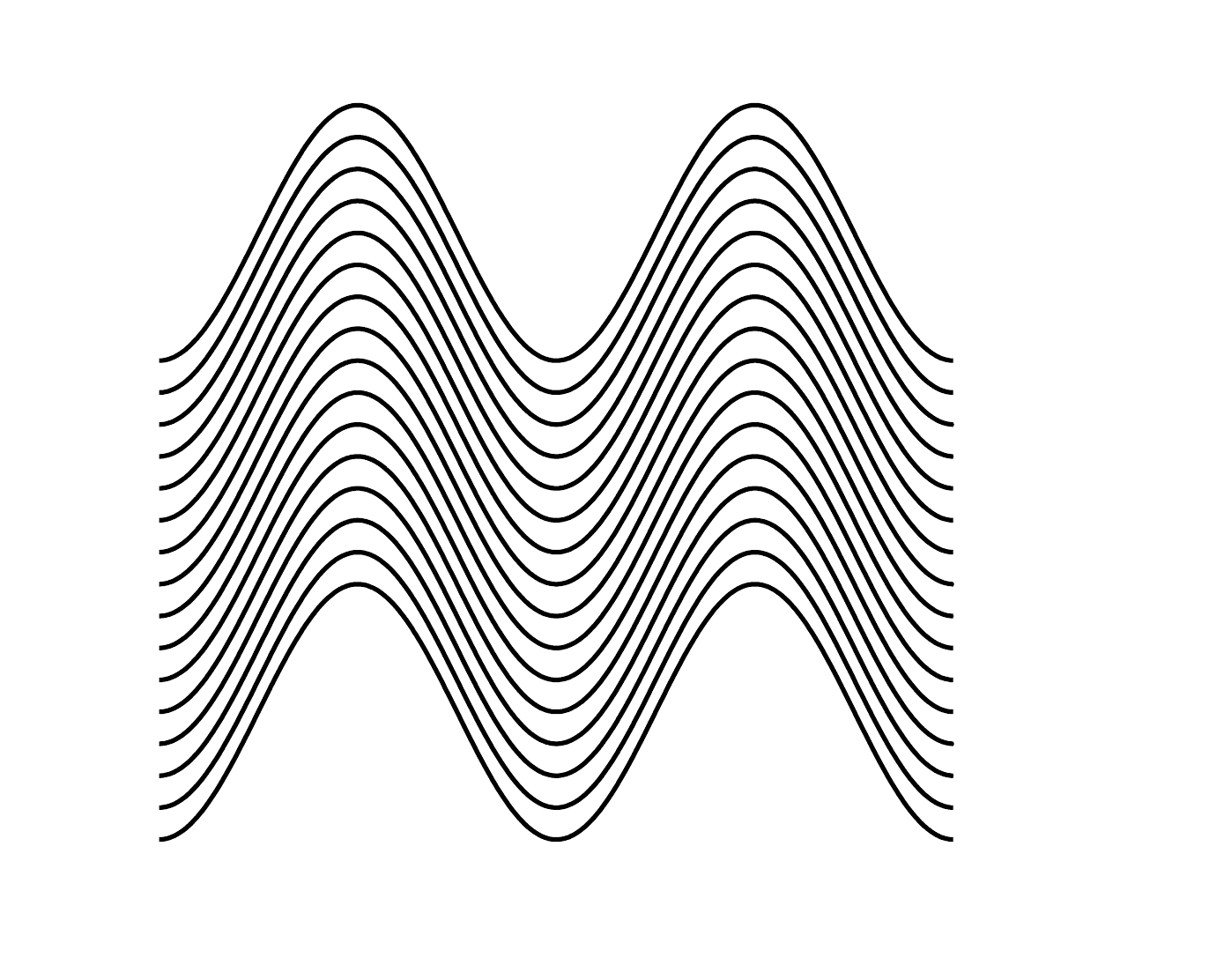}} &
\subfigure[Small phase difference]{\includegraphics[height=6cm]{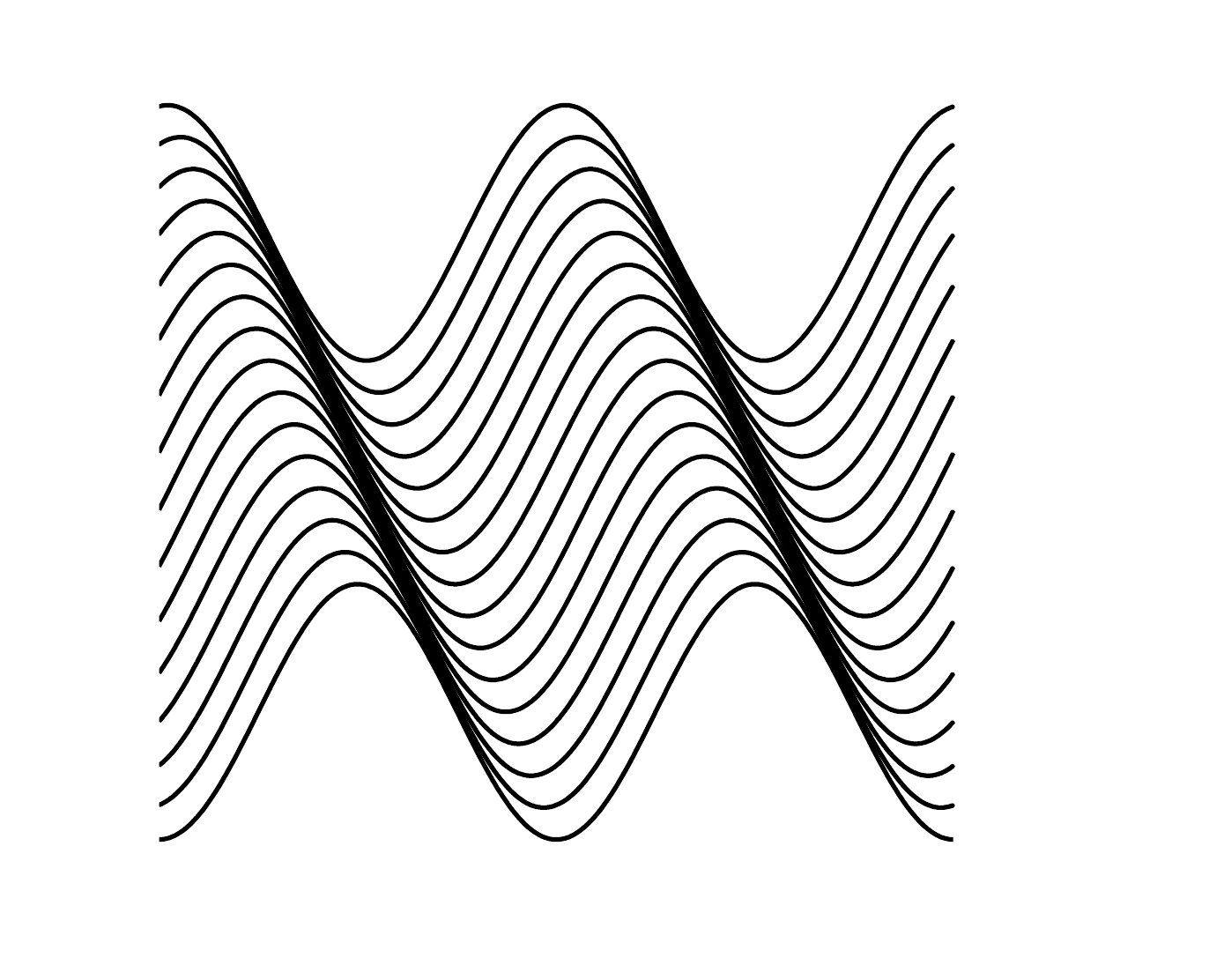}}
\end{tabular}
\caption{A small phase-difference between trajectories of neighboring Lagrangian points with identical and symmetric individual trajectories can lead to symmetry breaking and the appearance of a shock-like structure. (a) Harmonic in-phase beating. (b) Same harmonic motion with a linearly spatially varying phase.}\label{fig:shift}
\end{center}
\end{figure}

This small phase difference results in a global wave pattern at the organism level propagating from the south pole ($\theta=180^\circ$) to the north pole ($\theta=0^\circ$) as clearly seen on Fig.~\ref{fig:movie_reference}. We illustrate schematically in Fig.~\ref{fig:shift}  this symmetry-breaking mechanism through the collective behavior of individual Lagrangian points. With purely identical and symmetric motions of equal amplitude of neighboring Lagrangian points, the introduction of a small spatial phase  shift breaks the symmetry and generates a shock-like dynamics similar to the recovery stroke obtained numerically and shown in  Fig.~\ref{fig:theta_plot_reference}. 
For biological cells, the  biophysical origin of the observed phase-locking between neighboring cilia and the generation of metachronal waves  is still a matter of investigation \cite{gueron1997,vilfan2006,lenz2006,guirao2007,niedermayer2008}. Our results show  that a wave-like deformation of the surface at the whole-organism level is actually an optimal for the swimming efficiency. 

Finally, we note that the efficiency obtained through our optimization procedure is much larger than that of typical swimming microorganisms. Our optimal unconstrained stroke is found to be associated with very large displacements of material points on the spherical surface, with a maximum amplitude of $\Theta_\textrm{max}\approx 52.6^\circ$ (meaning that there exists a Lagrangian point for which  $\vartheta(\theta_0,t)-\theta_0$ varies between $-52.6^\circ$ and $+52.6^\circ$),  corresponding to a {linear tangential displacement equivalent to $90\%$ of a diameter}. This {stroke} of large amplitude is however not realistic  for real ciliated organisms as the length of the cilia (and therefore the maximum distance covered by its tip) is typically several times smaller than the size of the organism itself \cite{brennen1977}. In the following section our optimization approach  is adapted to constrain the maximum displacement amplitude of each surface point, and we study the maximum achievable swimming efficiency with a given maximum individual tangential displacement. The optimal unconstrained stroke obtained above
  will serve as a reference for comparison and discussion in the rest of the paper.

%%%%%%%%%%%%%%%%%%%%%%%%%%%%
% NEW SECTION: CONSTRAINED CASE
%%%%%%%%%%%%%%%%%%%%%%%%%%%%

\section{Stroke optimization with constrained surface displacements}
\label{sec:constr}
\subsection{Applying a constraint on the maximum displacement}
The optimization algorithm described in Sec. \ref{sec:unconstr} can be adapted to include a constraint on the maximum displacement of individual points on the surface, by modifying the function to be maximized as
\begin{equation}\label{eq:costfn}
J=\eta[\xi]-\int_{-1}^1H(T[\xi](x_0)-c)\dd x_0,
\end{equation}
where $\eta[\xi]$ is the efficiency as defined in Eq.~\eqref{eq:efficiency}, $T[\xi]$ is a measure of the amplitude of the displacement of individual material points for the stroke $\xi(x_0,t)$, and $c$ is a dimensionless threshold parameter (a smaller $c$ corresponding to a stricter constraint). $H$ is defined as
\begin{equation}\label{eq:penalfn}
H(u)=\Lambda\left[1+\tanh\left(\frac{u}{\epsilon}\right)\right]u^2.
\end{equation}
This form of $H$, when  $\Lambda$ is large and $\epsilon$ is  small,
introduces a numerical penalization in the cost function only when the displacement measure $T$ is greater than the threshold value. Values of $\Lambda=10^4$ and $\epsilon=10^{-3}$ were typically used in our numerical calculations, with little effect on the final result. 

Physically, the constraint $T[\xi]=\Theta[\xi]$ should ideally be applied, with 
\begin{equation}\label{eq:max_disp}
\Theta[\xi](x_0)=\frac{\theta_\textrm{max}(\theta_0)-\theta_\textrm{min}(\theta_0)}{2}=\frac{\textrm{max}\,_{t}\left[\cos^{-1}\left(\xi(x_0,t)\right)\right]-\textrm{min}\,_{t}\left[\cos^{-1}\left(\xi(x_0,t)\right)\right]}{2},
\end{equation}
which is the actual displacement amplitude of an individual point. However, the strong non-linearity of this measure is not appropriate for the computation of a gradient in functional space as presented in Sec. \ref{sec:unconstr}. An alternative measure of the displacement is
\begin{equation}\label{eq:disp_0eas}
T[\xi](x_0)=\mean{(\xi-\mean{\xi})^2}=\frac{1}{2\pi}\int_0^{2\pi}\left[\xi(x_0,t)-\frac{1}{2\pi}\int_0^{2\pi}\xi(x_0,t')\dd t'\right]^2\dd t,
\end{equation}
which is the variance of the displacement along the vertical axis. 
Note that in using Eq.~\eqref{eq:disp_0eas},  we do not strictly enforce that the maximum displacement along the surface should be less than a given threshold. In addition, choosing  Eq.~\eqref{eq:disp_0eas} instead of Eq.~\eqref{eq:max_disp}  introduces a difference between points located near the poles (a small change in $\theta$ there corresponds to a much smaller vertical displacement  compared to that obtained with the same angular displacement at the equator). This is however not an issue as Eq.~\eqref{eq:disp_0eas} will still penalize large amplitude strokes and the constraint applied using Eq.~\eqref{eq:disp_0eas} will be stronger near the equator, where the displacements were observed the largest in the unconstrained problem (Fig.~\ref{fig:theta_plot_reference}).   In addition, a posteriori, an implicit relationship between $c$ and the actual maximum displacement can be obtained, and $c$ can thus be used as a tuning parameter to constrain the maximum amplitude $\Theta_\textrm{max}$.

\begin{figure}
\begin{center}
\includegraphics[width=9cm]{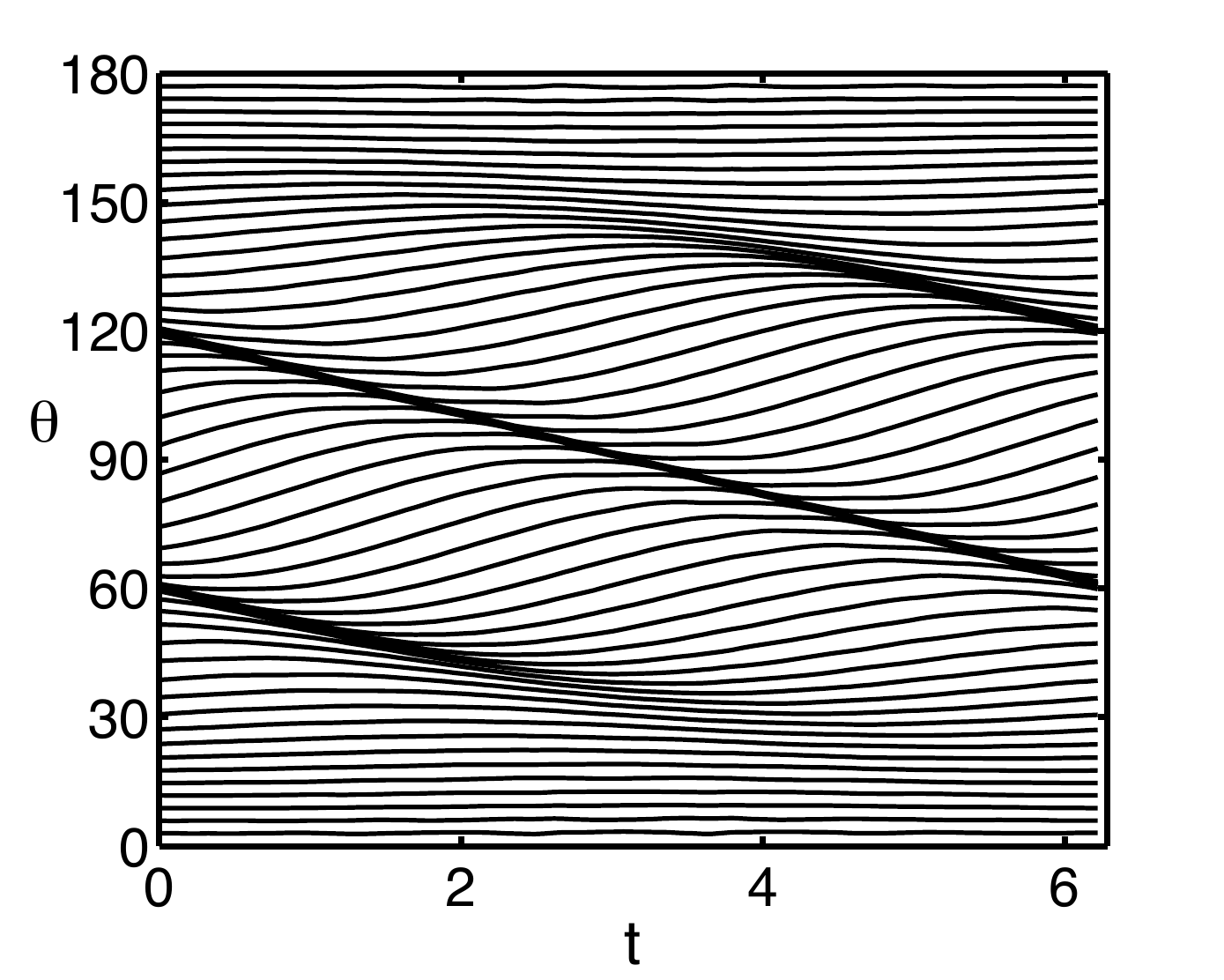}
\caption{(color online) Same as Fig.~\ref{fig:theta_plot_reference} in a
  constrained-displacement case. Here, the maximum displacement is $\Theta_\textrm{max}\approx10.6^\circ$, {the mean swimming velocity is $\mean{U}\approx0.058$} and the swimming efficiency is $\eta\approx 6.2\%$.}\label{fig:theta_plot_constrained}
\end{center}
\end{figure}

Following the same approach as in Sec. \ref{sec:unconstr}, we consider a small perturbation $\dxi$ of a reference swimming stroke $\txi$. The resulting change in the cost function $J[\xi]$ is obtained as
\begin{equation}\label{eq:grad_constr}
\delta J=\int_0^{2\pi}\int_{-1}^1\mathcal{F}[\txi](x_0,t)\dxi(x_0,t)\dd x_0 \dd t,
\end{equation}
where the modified gradient $\mathcal{F}$ is now
\begin{equation}\label{eq:grad_q_bis}
\mathcal{F}[\txi]=F[\txi]-\frac{\Lambda}{\pi}\left(\txi-\mean{\txi}\right)H'\left[T[\txi]-c\right],
\end{equation} 
and the corrected gradient $\mathcal{G}[\tpsi]$ of the cost function with respect to $\psi$ is then obtained from $\mathcal{F}$ as in Eq.~\eqref{eq:grad_psi}.

\begin{figure}
\begin{center}
\includegraphics[width=15cm]{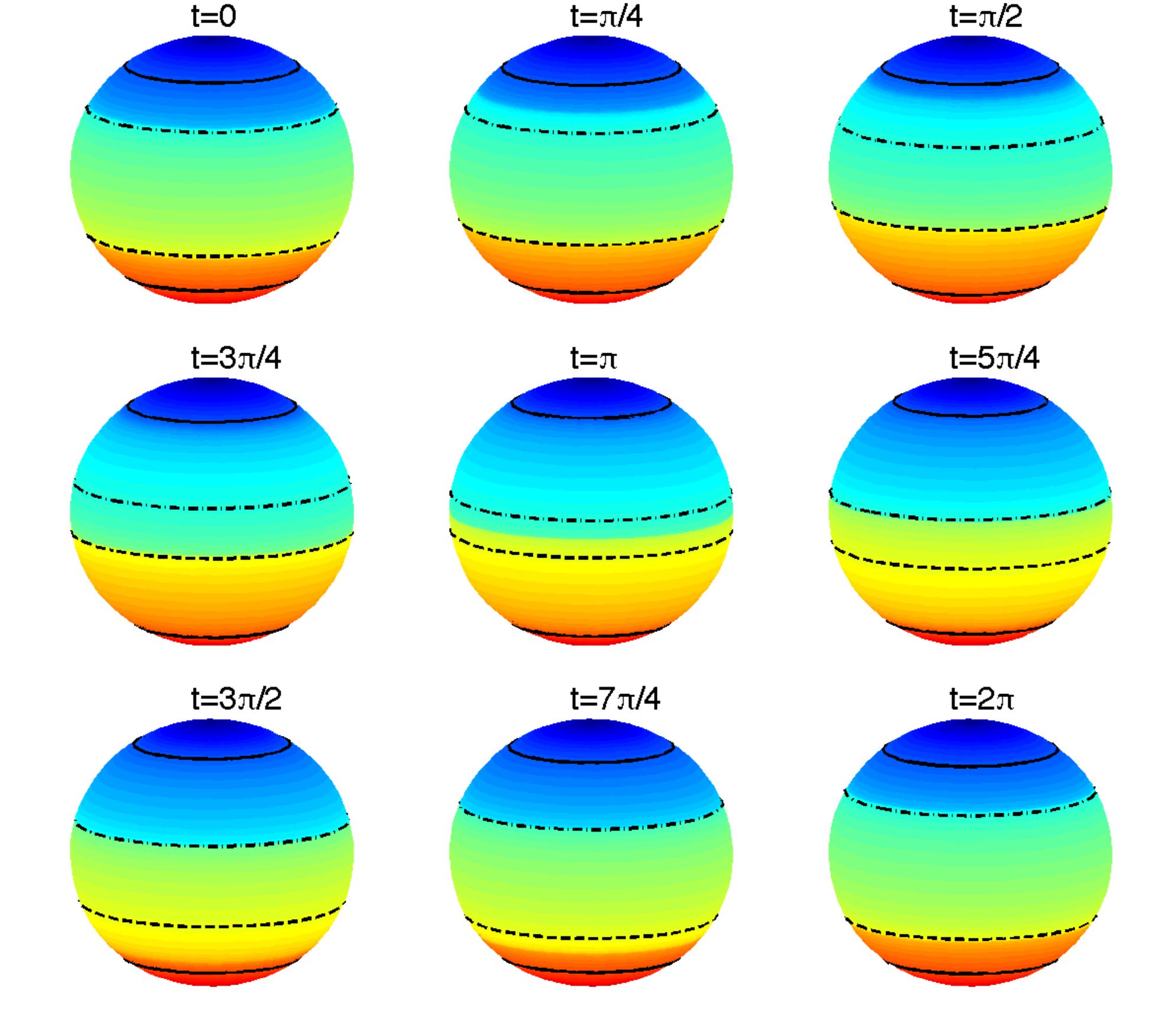}
\caption{Same as Fig.~\ref{fig:movie_reference} in the constrained
  case. The example illustrated here has a maximum displacement amplitude
  $\Theta_\textrm{max}\approx10.6^\circ$, {a mean swimming velocity $\mean{U}\approx0.058$} and a swimming efficiency $\eta\approx 6.2\%$.}\label{fig:movie_constrained}
\end{center}
\end{figure}

%%%%% 
\subsection{Results}

\begin{figure}
\begin{center}
\begin{tabular}{c}
\includegraphics[width=12cm]{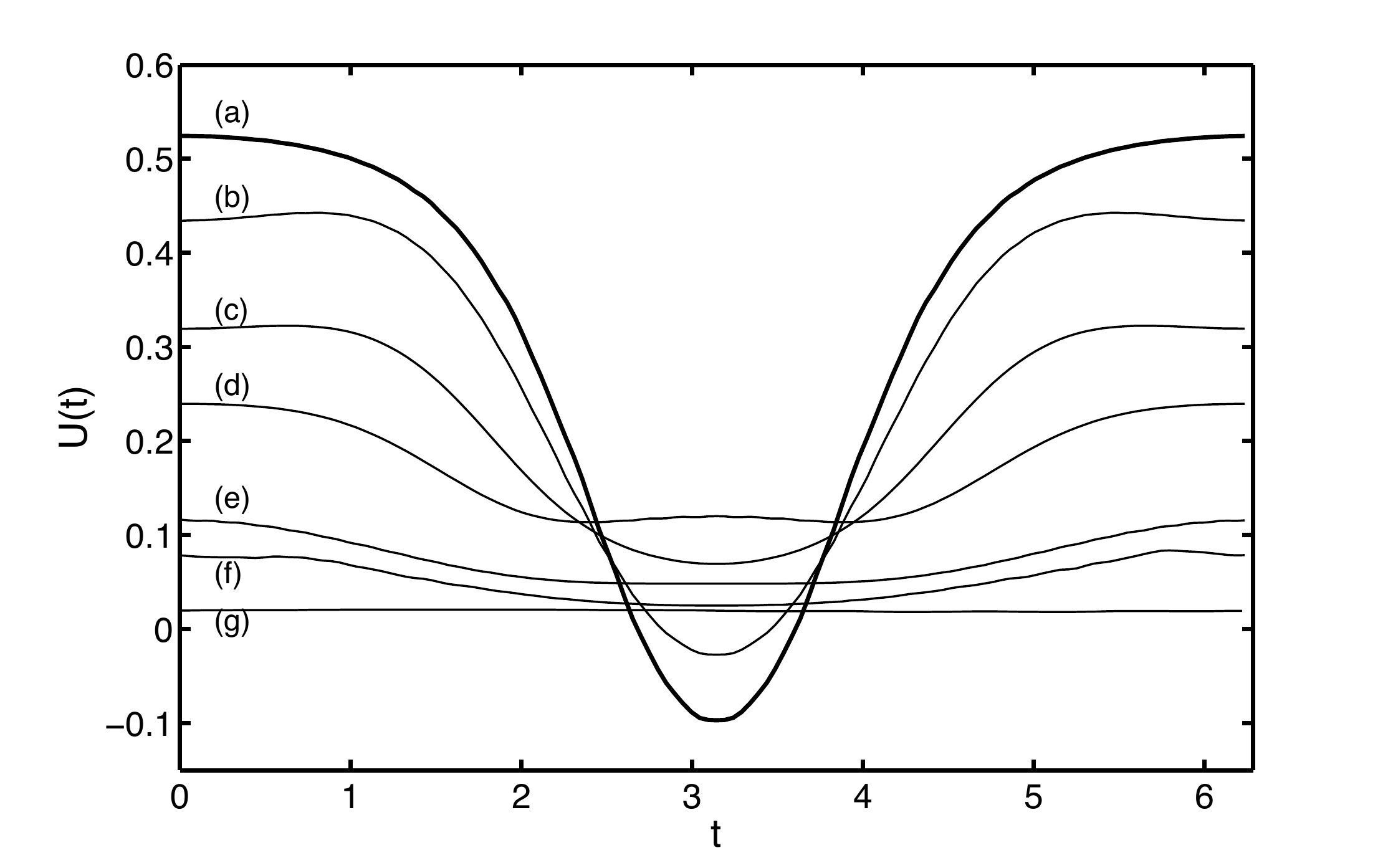} \\
\begin{tabular}{|c|ccccccc|}
\hline
Case & (a) & (b) & (c) & (d) & (e) & (f) & (g)\\
\hline
$\Theta_\textrm{max}$ & $52.6^\circ$ & $45.1^\circ$ & $34.2^\circ$ & $26.9^\circ$ & $14.4^\circ$ & $11.0^\circ$ &$4.26^\circ$ \\
\hline
$\eta$ (\%) &  $22.2$ &$21.0$ &$17.9$ &$15.1$ &$8.33$ &$6.81$ & $2.61$ \\
\hline
$\mean{U}$ & $0.33$&   $0.29$&   $0.22$&   $0.17$&   $0.075$&   $0.051$& $0.02$\\
\hline
\end{tabular}
\end{tabular}
\caption{{Variations of the swimming velocity over a period for optimal strokes with different maximum displacements. The unconstrained optimal stroke is represented by a thick black line (a). The values of the maximum displacement, the swimming efficiency and the mean swimming velocity are also given for each case}.}\label{fig:swimev}
\end{center}
\end{figure}

A number of simulations were performed using a range of values of $c$, each with various choices for  both $\Lambda$ and $\epsilon$. As expected, we observe that introducing the penalization and reducing the value of $c$ does indeed reduce the maximum displacement amplitude of individual Lagrangian points. This decrease in $\Theta_\textrm{max}$ is found to systematically be associated with a decrease in the  swimming efficiency. The swimming stroke obtained under constrained maximum displacements presents the same structure as the unconstrained case (see Figs.~\ref{fig:theta_plot_constrained} and \ref{fig:movie_constrained}).  The $(\theta,t)$ domain can be divided into two strokes, an effective stroke where the surface moves downward while stretching, and a recovery stroke, corresponding to the upward motion of a compressed surface. The maximum amplitude of displacement is however smaller, as is the propagation velocity of the shock structure corresponding to the recovery stroke (Fig.~\ref{fig:theta_plot_constrained}). {Because of this smaller shock velocity, the recovery and effective strokes co-exist over the entire swimming period on different regions of the swimmer's surface, resulting in a smoothing of the swimming velocity who becomes close to a constant when the constraint on $\Theta_\textrm{max}$ becomes tighter (Fig.~\ref{fig:swimev}).}

\begin{figure}
\begin{center}
\includegraphics[width=8.1cm]{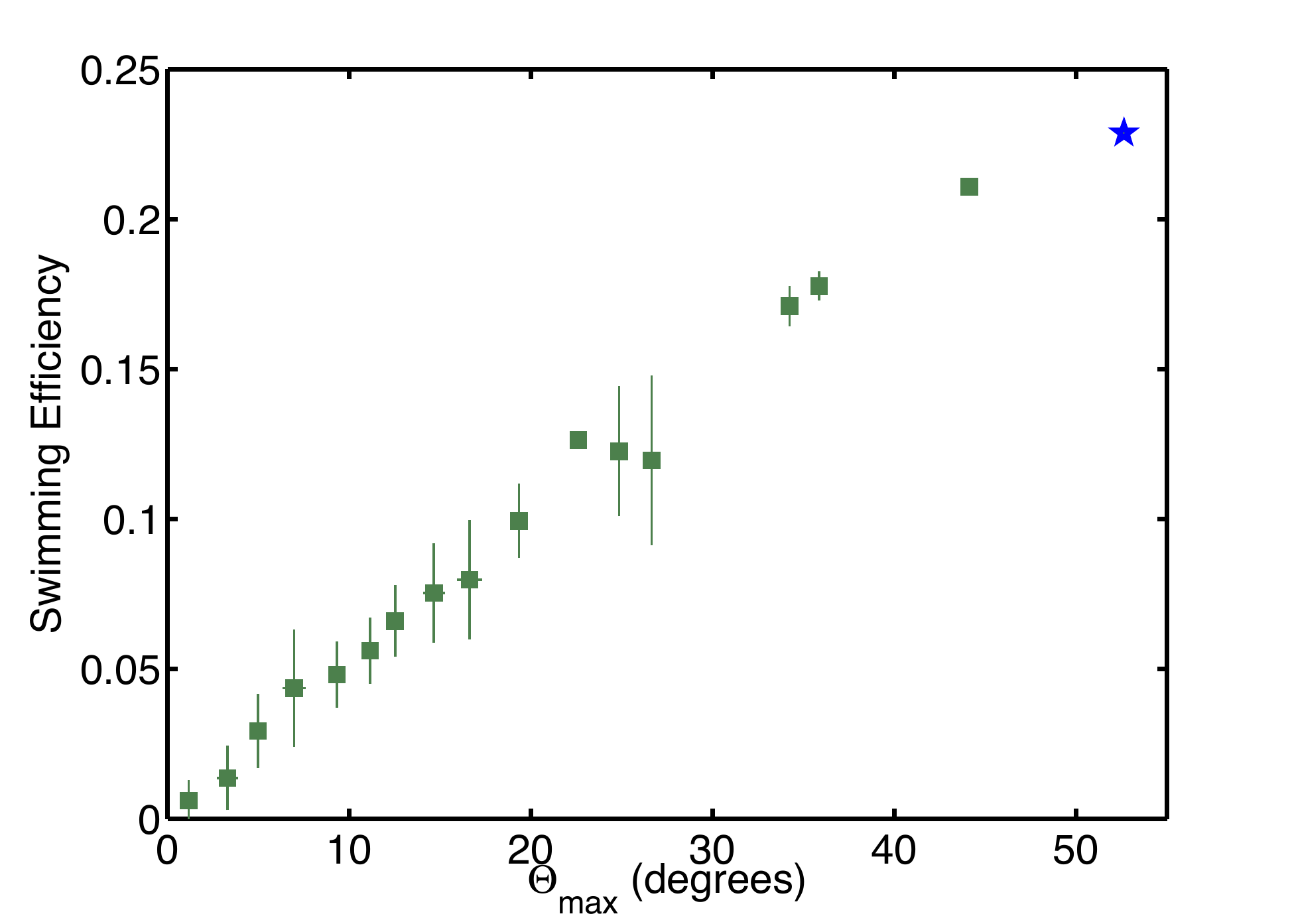}
\includegraphics[width=8.1cm]{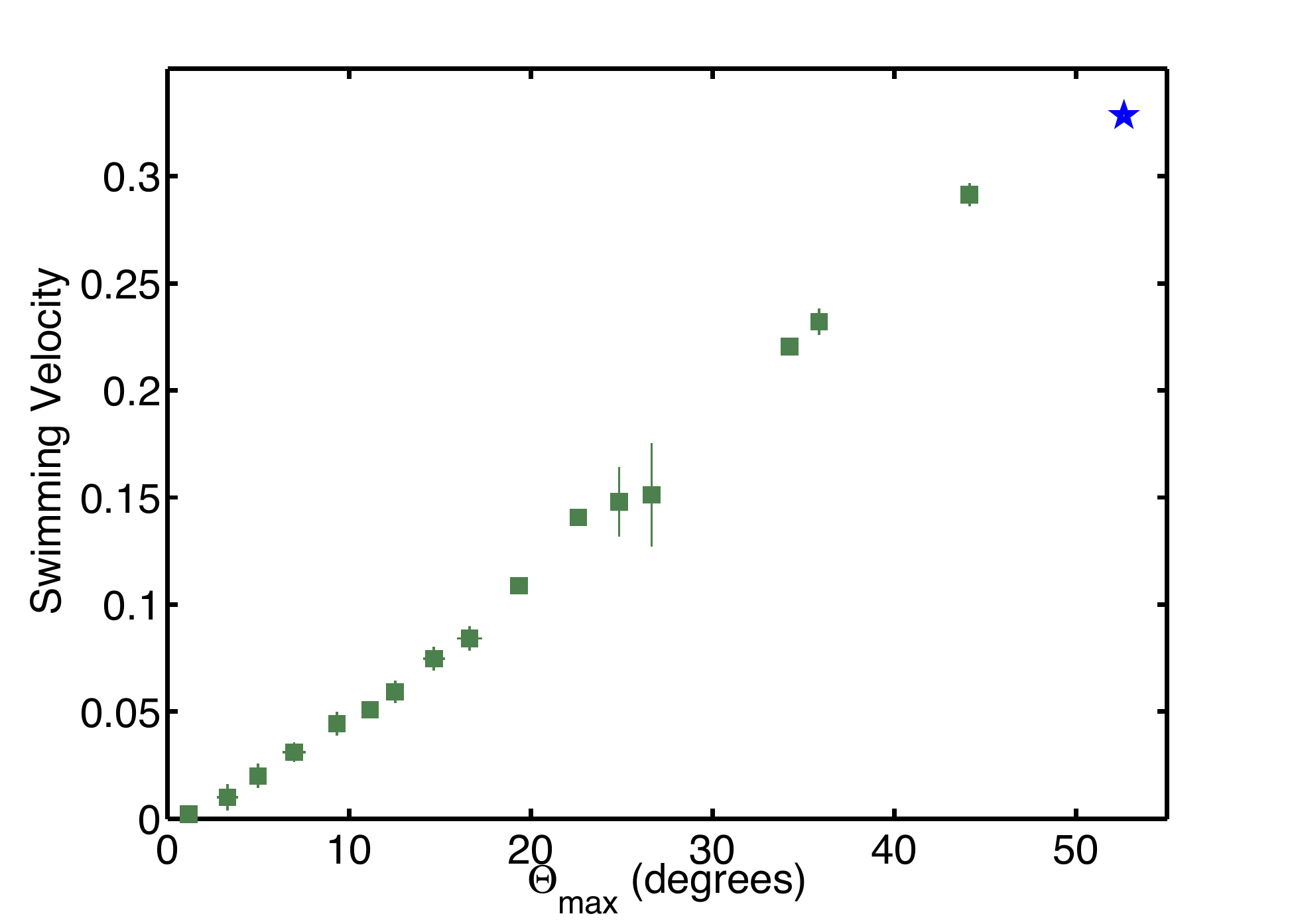}
\caption{(color online) Left: Swimming efficiency, $\eta$,  vs. the maximum amplitude in angular displacement of the stroke, $\Theta_\textrm{max}$. The results of multiple runs (with different initial conditions or penalizing function parameters) have been regrouped into bins. Horizontal and vertical error bars represent {the standard deviation} in displacement and efficiency for each point.  {The blue star corresponds to the optimal unconstrained swimming stroke of Sec. \ref{sec:unconstr}.}
Right: Swimming velocity vs.  maximum angular displacement.
}\label{fig:eff_th}
\label{fig:swim_vel}
\end{center}
\end{figure}

For each simulation, the maximum amplitude $\Theta_\textrm{max}=\textrm{max}(\Theta[\xi](x_0))$ can be computed together with the swimming efficiency $\eta$. A clear monotonic relationship between $\eta$ and $\Theta_\textrm{max}$ is obtained, as shown in Fig.~\ref{fig:eff_th} (left). The dispersion around the main trend is small and due to some numerical runs converging to local minima. The right-most point of the curve corresponds to the unconstrained optimization presented in Sec.~\ref{sec:unconstr}.
Similarly, the swimming velocity is found to be an increasing function of the maximum tangential displacement in the optimal stroke (Fig.~\ref{fig:swim_vel}, right). It is maximum for the unconstrained stroke ($\Theta_\textrm{max}\approx52.6^\circ$) with a dimensionless speed of $U\approx 0.33$, corresponding to about one body length per period, and is reduced to less than half a body length per period when the displacement amplitude is constrained below $\approx 30^\circ$.

%%%%%%%%%%%%%%%%%%%%%%
% NEW SECTION: SOLUTION ANALYTIQUE
%%%%%%%%%%%%%%%%%%%%%%

\section{Theoretical upper bound on the swimming efficiency of a
  squirmer}
  \label{sec:ansatz}

Using the numerical optimization approach described above, we have obtained an optimal swimming stroke with an efficiency of $\approx 22.2\%$. We have further shown  that constraining the maximum displacement of the surface reduces the swimming efficiency continuously from that maximum value. In this section, we are addressing the  question of whether the numerical result of about $22.2\%$ is the maximum achievable efficiency for a periodic swimming stroke. We show below that, in fact,   the theoretical upper bound of $50\%$ can be reached asymptotically by a singular  stroke.
  
\subsection{Bound on the swimming efficiency and conditions of
  equality}

 {The Cauchy-Schwartz inequality states that} $\mean{\alpha_1^2}\geq\mean{\alpha_1}^2$ and the equality can only be reached when $\alpha_1(t)=\mean{\alpha_1}$ is a constant. From the definition of the swimming efficiency $\eta$ in   Eq.~\eqref{eq:efficiency}, we therefore obtain that for any swimming stroke $\xi$
  \begin{equation}\label{eq:etabound}
  \eta[\xi]\leq\frac{1}{2},
  \end{equation}
  and this upper bound can only be reached for the particular case:
  \begin{equation}\label{eq:equalitycondition}
  \alpha_1(t)=\mean{\alpha_1}\quad \textrm{and}\quad\alpha_n(t)=0\,\,\textrm{for all  }n\geq 2.
  \end{equation}
  In this case, $u_\theta(\theta,t)\sim\sin\theta$ and the vertical surface velocity is constant and uniform. This stroke corresponds to the so-called ``treadmilling" swimmer \cite{leshansky2007} which is not time-periodic  as the poles are permanent source and sink of material surface points, which move continuously from one pole to the other.
  
 However, the treadmill swimmer provides some important insight about strokes maximizing the efficiency: $\alpha_1$ must be dominant over the other modes to minimize the energy consumption. This observation is consistent with the relative mode amplitudes $\alpha_n(t)$ observed in the result of the optimization procedure (Fig.~\ref{fig:alpha}). 
To guarantee the stroke periodicity, {at least} one of the two conditions in Eq.~\eqref{eq:equalitycondition} must be violated. In the following, we build an analytical ansatz that achieves $\eta=1/2$ asymptotically based on the observation that the conditions of Eq.~\eqref{eq:equalitycondition} need only to be satisfied on most of the period, the interval where they are violated being asymptotically of zero measure.
  
\subsection{Building the analytical ansatz}
The analytical ansatz is the superposition of two parts, the effective stroke or outer solution which covers most of the $(x,t)$-plane, and the recovery stroke or inner solution, which takes the form of a shock and enforces periodicity of the surface displacement.

\subsubsection{Outer solution}
The outer solution satisfies the optimality conditions of Eq.~\eqref{eq:equalitycondition} exactly. The Lagrangian equation of motion of the surface points is obtained from Eqs. \eqref{eq:motioneq_x} and \eqref{eq:alphndef} as
\begin{equation}
\pard{\xi}{t}=\gamma(x^2-1)\quad \textrm{with} \quad \gamma=\frac{3}{2}\alpha_1,
\end{equation}
which can be integrated as
\begin{equation}\label{eq:out_sol}
\xi(x_0,t)=\tanh\left(\log\sqrt{\frac{1+x_0}{1-x_0}}-\gamma t\right).
\end{equation}
Using the change of variables $x=\tanh y$ and $\xi=\tanh \chi$, Eqs. \eqref{eq:alphndef} and \eqref{eq:out_sol} become
\begin{align}
\label{eq_0otion_2}
\pard{\chi}{t}(y_0,t)=-&\sum_{n=1}^\infty\frac{2n+1}{n(n+1)}\alpha_n(t)L_n'\left[\tanh(\chi)\right],\qquad -\infty<y<\infty,\quad -\pi\leq t\leq \pi,\\
\textrm{Outer solution:    } &\chi(y_0,t)=y_0-\gamma t, -\infty<y_0<\infty.\label{outer_sol_y}
\end{align}
The trajectories of the outer solution are straight lines in the $(\chi,t)$-plane and are not periodic since
\begin{equation}
\chi(y_0,\pi)-\chi(y_0,-\pi)=-2\gamma\pi.
\end{equation}

\begin{figure}
\begin{center}
\begin{tabular}{c}
\subfigure[Trajectories $\theta=\vartheta(\theta_0,t)$ and shock position]{\includegraphics[width=13cm]{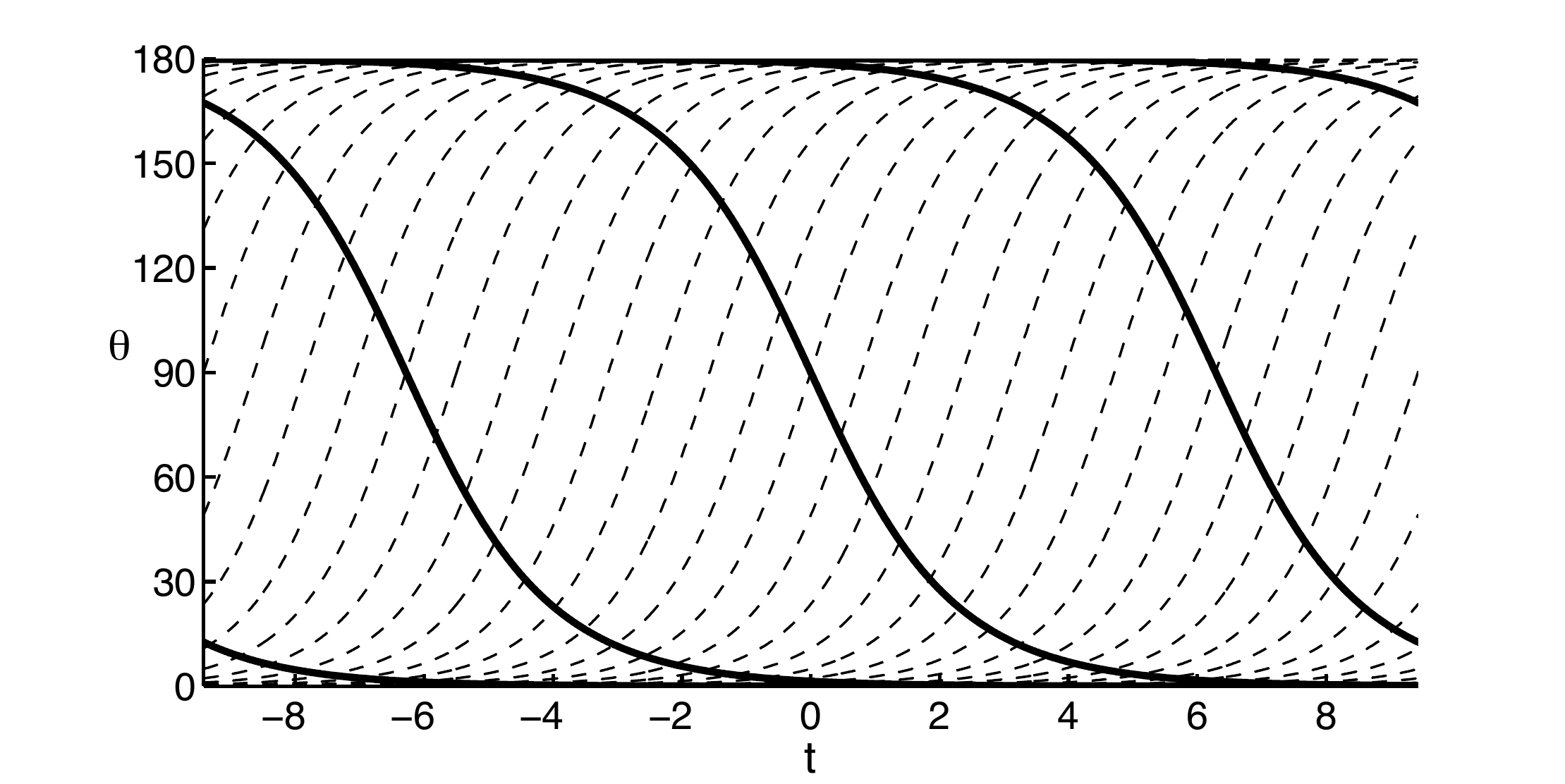}} \\
\subfigure[Trajectories $z=\varsigma(z_0,\tau)$ and shock position]{\includegraphics[width=13cm]{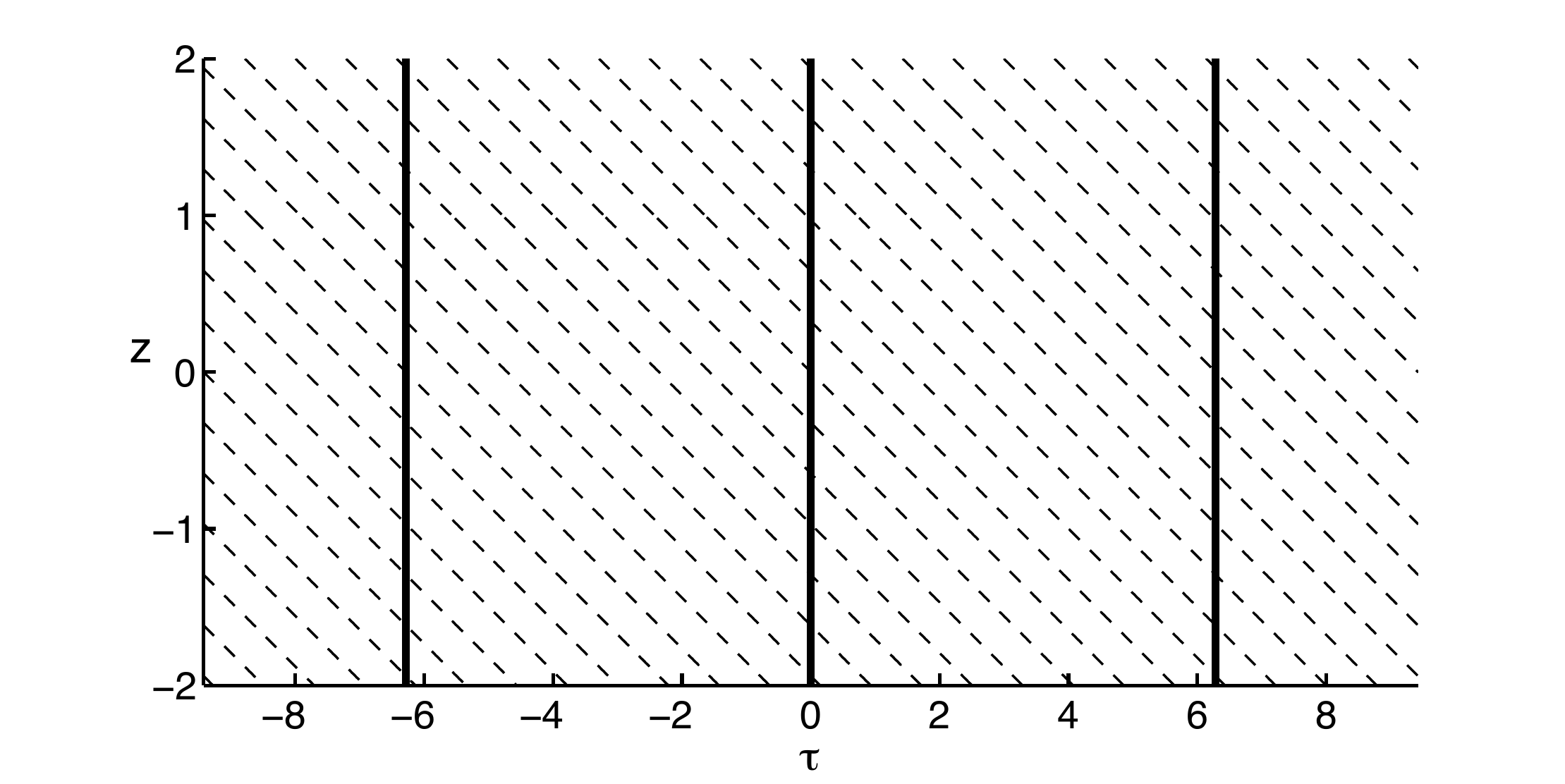}} 
\end{tabular}
\caption{The analytical ansatz is decomposed into an outer solution (effective stroke) in which only the swimming mode is present and an inner solution (recovery stroke) of small width  which enforces periodicity in time. The trajectories corresponding to the outer solution are shown in dashed lines, and thick solid lines show the position of the shocks and inner solution trajectories (Eq.~\ref{eq:shock_eq}). Here, $\gamma=1$ and $\beta=0.7$, and three swimming periods are shown. }\label{fig:schema_ansatz}
\end{center}
\end{figure}

\subsubsection{Shocks and inner solution}
To construct a time-periodic solution, thin shock-like structures are super imposed to this outer solution (Fig.~\ref{fig:schema_ansatz}). We look for shock equations of the form:
\begin{equation}\label{eq:shock_eq}
x_{S_k}=\tanh[\beta(t+2k\pi)]\quad \textrm{or}\quad y_{S_k}=\beta(t+2k\pi).
\end{equation}

In the $(y,t)$-plane, a Lagrangian surface particle will follow a straight line of slope $-\gamma$, until it hits one of the shocks (Fig.~\ref{fig:schema_ansatz}). The velocity inside the shock must be designed in such a way that the Lagrangian point will reemerge from the shock when $\Delta y =2\pi\gamma$. The following change of variables, $z=y$ and $\tau=t-y/\beta$ stretches the horizontal coordinate and positions the shocks as vertical lines $\tau=2k\pi$ in the $(z,\tau)$-plane (see Fig.~\ref{fig:schema_ansatz}b).  The trajectories of Lagrangian particles $y=\chi(y_0,t)$ are now rewritten as $z=\varsigma(z_0,t)$, with
\begin{equation}
\pard{\chi}{t}=\pard{\varsigma}{\tau}\left(1+\frac{1}{\beta}\pard{\varsigma}{\tau}\right)^{-1}\cdot
\end{equation}
In $(z,\tau)$ coordinates, the outer solution trajectories can be expressed as 
\begin{equation}
\varsigma(z_0,\tau)=z_0-\frac{\gamma\tau}{1+{\gamma}/{\beta}}\cdot
\end{equation}
The Eulerian velocity $w(z,\tau)$ in $(z,\tau)$-coordinates is now a function of $\tau$ only (Fig.~\ref{fig:schema_ansatz}) and $w(\tau)=-\gamma\beta/(\gamma+\beta)$ everywhere, except in thin regions near $\tau=2k\pi$.  Considering the Gaussian approximation for $\delta_\sigma(\tau)$,
\begin{equation}
\delta_\sigma(\tau)=\frac{\ee^{-{\tau^2}/{\sigma^2}}}{\sigma\sqrt{\pi}},
\end{equation}
the solutions of
\begin{equation}
\pard{\varsigma}{\tau}=w(\tau)\quad \textrm{with}\quad w(\tau)=\frac{\gamma\beta}{\gamma+\beta}(2\pi\sum_{k=-\infty}^\infty\delta_\sigma(\tau+2k\pi)-1)
\end{equation}
are periodic and equal to the outer solution outside thin regions of characteristic width $\sigma$. Changing back to variables $(y,t)$ and $(x,t)$, the ansatz is finally obtained as
\begin{equation}
\label{eq:ansatz_sol_2}
\pard{\xi}{t}=u(x,t)=(1-x^2)\left[-\gamma+(\beta+\gamma)\frac{\displaystyle\sum_{k=-\infty}^\infty\delta_\sigma\left(t-\frac{\tanh^{-1}x}{\beta}+2k\pi\right)}{\displaystyle\frac{\beta}{2\pi\gamma}+\sum_{k=-\infty}^\infty\delta_\sigma\left(t-\frac{\tanh^{-1}x}{\beta}+2k\pi\right)}\right]\cdot
\end{equation}

For given values of $\beta$, $\sigma$ and $\gamma$, the efficiency of the corresponding periodic stroke can be computed numerically by projecting $u(x,t)$ using Eq.~\eqref{eq:alphn_comp}. The infinite sums in Eq.~\eqref{eq:ansatz_sol_2} can be easily truncated after a few terms as the contribution of larger values of $k$ to the term in brackets is limited to the vicinity of $\pm 1$. 

\subsection{Asymptotic convergence to $\eta=50\%$}
\begin{figure}
\begin{center}
\begin{tabular}{cc}
\includegraphics[width=8cm]{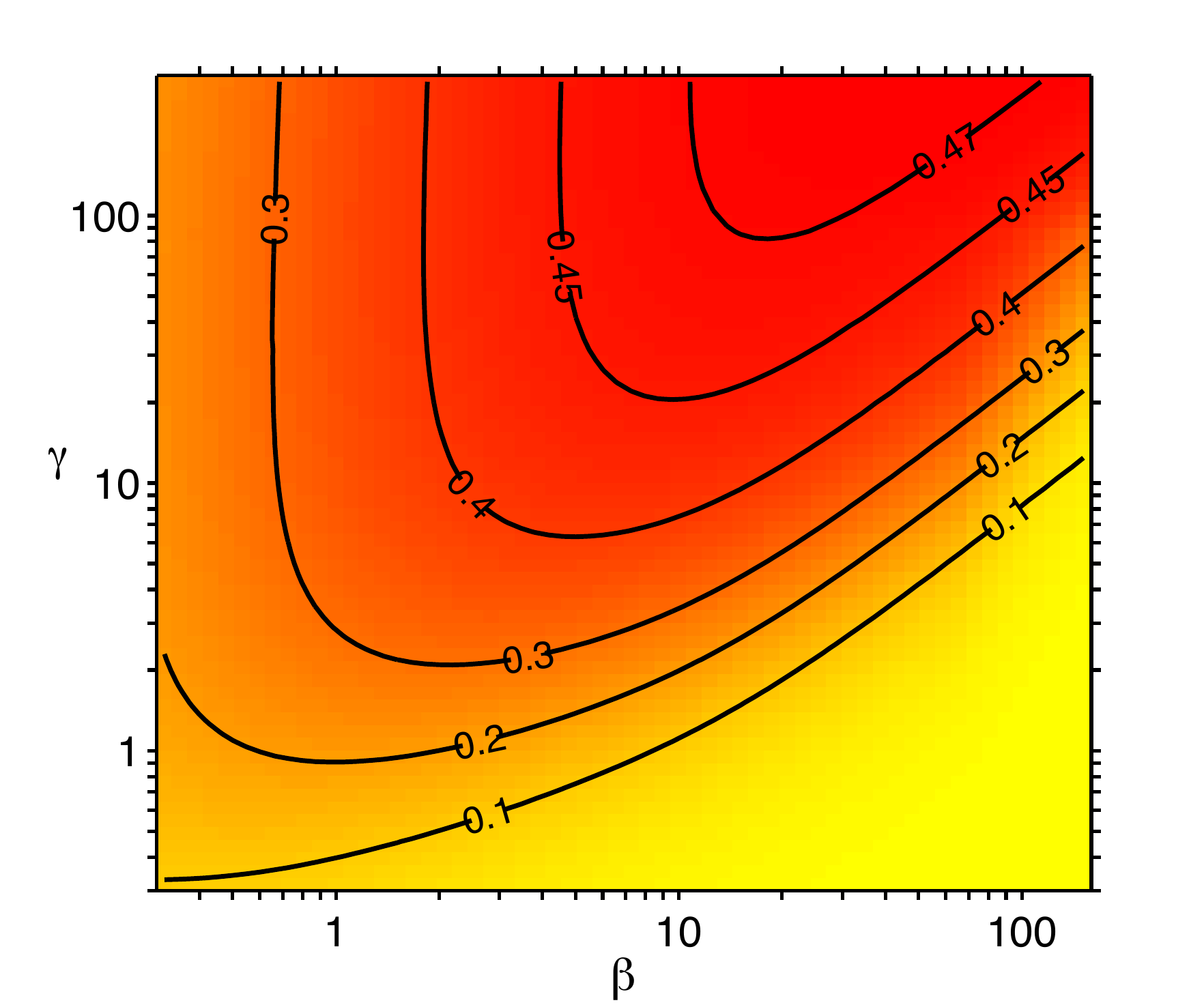}&
\includegraphics[width=8.2cm]{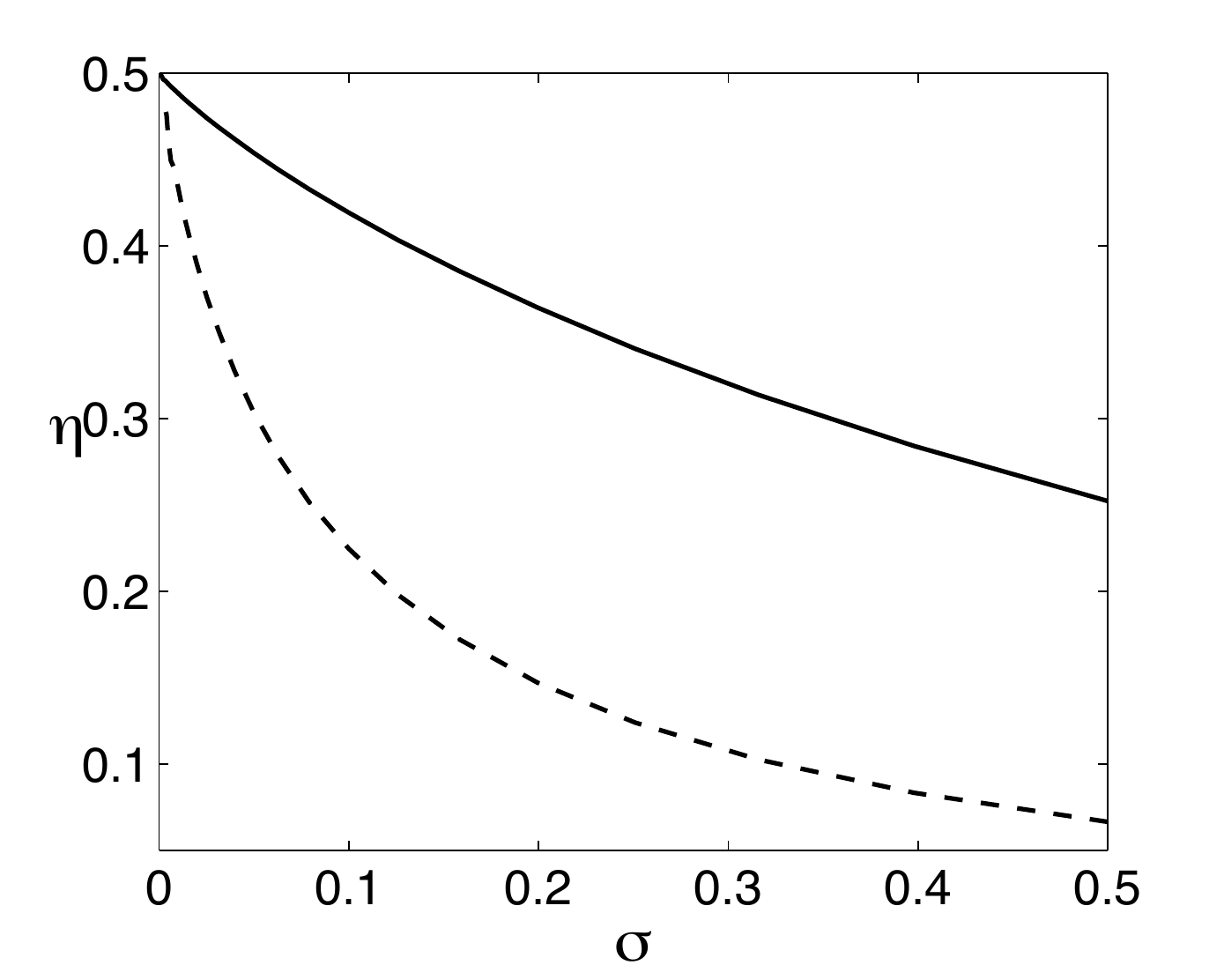}
\end{tabular}
\caption{(color online)  (Left) Variations of the swimming efficiency obtained for the analytical
  ansatz, for a sweep in the ($\beta,\gamma$) parameter space, and a shock width parameter
of $\sigma=0.05$; lines display iso-values of $\eta$.  (Right) Evolution with $\sigma$ of the efficiency of the analytical ansatz for $\gamma=1000$ (solid) and $\gamma=200$ (dashed); in both cases, $\beta=800$.}\label{fig:span_eff}
\end{center}
\end{figure}

The solution of Eq.~\eqref{eq:ansatz_sol_2} satisfies the periodicity constraint and matches the (optimal) treadmilling swimming stroke over most of the domain. For a small value of the shock width $\sigma$, we numerically sweep the  parameter space $(\beta,\gamma)$, and display the map of the corresponding values of the swimming efficiency  in Fig.~\ref{fig:span_eff}. We observe that when $\beta$ and $\gamma$ become very large, values of the efficiency approaching $50\%$ can be reached. Figure \ref{fig:span_eff} also shows the variations of $\eta$ with $\sigma$ for large values of $\beta$ and $\gamma$. We thus obtain that the value of  $\eta=50\%$ can be  obtained in the asymptotic limit where $\sigma\rightarrow 0$ and $\beta,\gamma\rightarrow\infty$.

The first criterion ($\sigma\rightarrow 0$) is expected to occur as the shock must be narrow   for the measure of the domain where the outer solution does not hold to be small.  In addition, we  see that the shock velocity ($\beta$) and swimming velocity ($2\gamma/3$) must also go to infinity. This is the result of the non-dimensionalization of the equations using the stroke frequency. When $\gamma\rightarrow\infty$, the effective stroke accumulates an asymptotically infinite amount of surface at one pole between two successive recovery strokes; $\beta\rightarrow\infty$ corresponds to the recovery stroke taking as little time as possible to bring back this accumulated surface near the opposite pole, thereby to a time-periodic stroke. If the swimming velocity had been used for non-dimensionalization, the limit $\gamma\rightarrow\infty$ would have corresponded to an infinite spacing between two successive recovery strokes. The present asymptotic solution with $\beta,\gamma\rightarrow\infty$ therefore matches the treadmill swimmer except for a time-interval of measure zero.
In this singular (and thus, non-physical) limit, the entire surface of the sphere travels from one pole to the other during the effective stroke before the shock/recovery stroke redistributes the surface points, corresponding to a maximum displacement amplitude $\Theta_\textrm{max}$ is equal to $90^\circ$.

\subsection{Efficiency vs. maximum displacement -- Final optimization diagram}
  
We therefore established numerically that for any value of $\eta$ less than, but arbitrarily close to, $50\%$, one can find parameter values for $\beta$, $\gamma$ and $\sigma$ for which the stroke (Eq.~\ref{eq:ansatz_sol_2}) leads to  swimming with efficiency $\eta$. This upper bound is therefore reachable asymptotically but the corresponding stroke is singular. It is therefore not surprising that such a stroke, where all Lagrangian points accumulate at the south pole, could not be obtained through the numerical optimization procedure presented in Secs.~\ref{sec:unconstr} and \ref{sec:constr}, which is based on a Lagrangian description of the surface

\begin{figure}
\begin{center}
\includegraphics[width=15cm]{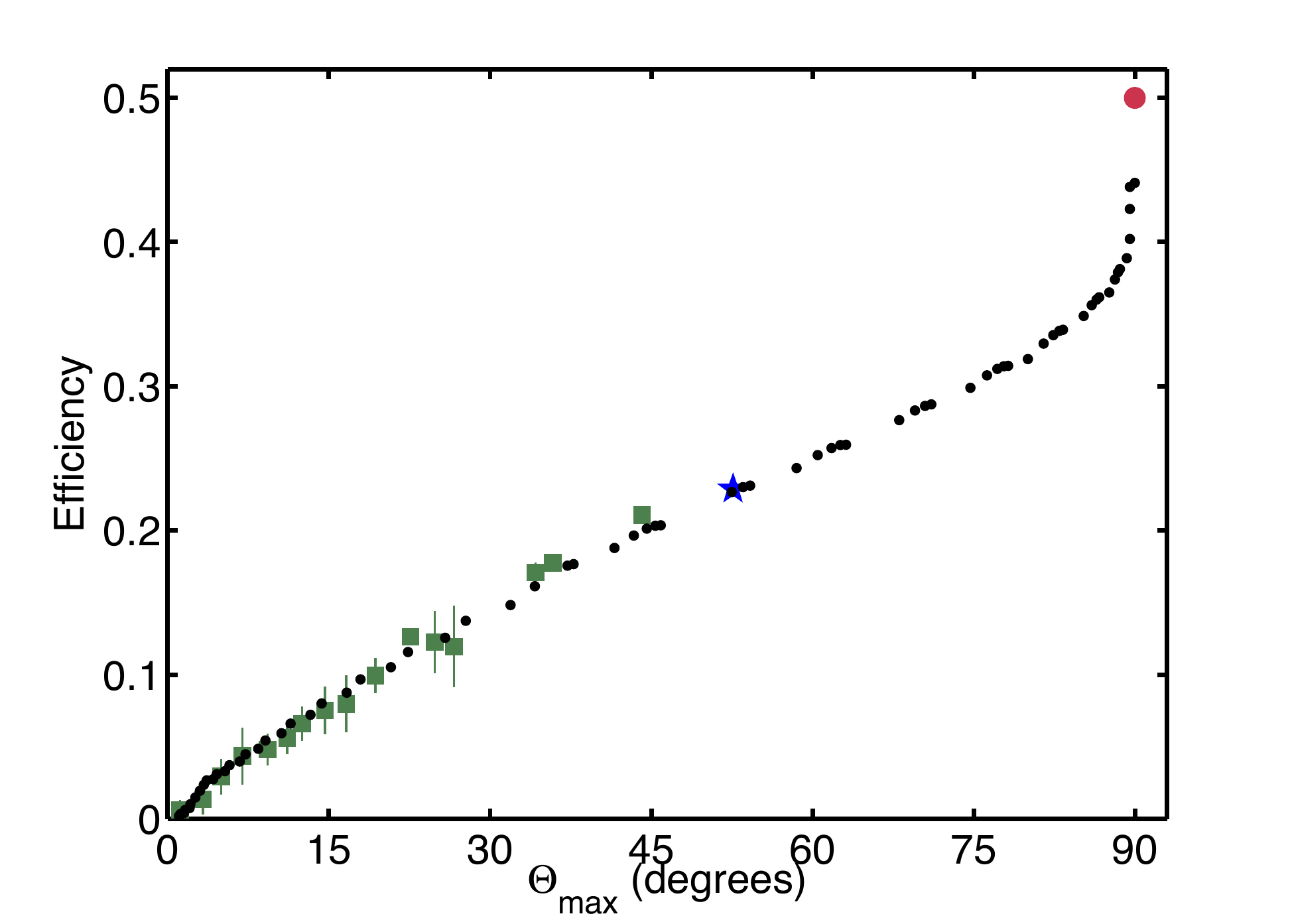}
\caption{(color online)  Final optimization diagram: Optimal efficiency of the swimming stroke, $\eta$, 
  as a function of the maximum Lagrangian angular displacement of the surface, $\Theta_\textrm{max}$. Green squares: result from numerical optimization (the
  vertical and horizontal error bars denote the variability from the
  initial conditions and the weight of the constraint); {blue star: unconstrained optimal swimming stroke with $\Theta_\textrm{max}\approx 52.6^\circ$ and $\eta\approx 22.2^\circ$; }black dots:  envelope of the results obtained   
  from the analytical ansatz for $\sigma=0.1$ and varying $\beta$ and $\gamma$; red dot:   asymptotic limit of
  the singular quasi-treadmill swimmer of efficiency $\eta=1/2$ and 
   maximum angular amplitude $\Theta_\textrm{max}=90^\circ$.}\label{fig:masterplot}
\end{center}
\end{figure}

The analytical approach can however also be used to confirm the results of our optimization approach for small values of the swimming efficiency. Equation \eqref{eq:ansatz_sol_2} defines a family of periodic strokes with three parameters $\beta$, $\gamma$ and $\sigma$. For each stroke, we can compute numerically its  efficiency together with the maximum displacement amplitude $\Theta_\textrm{max}$. {The envelope curve, or equivalently the maximum efficiency obtained for a given maximum displacement amplitude, can then be compared to the results of the optimization procedure from Secs.~\ref{sec:unconstr} and \ref{sec:constr}.} The results of this comparison are displayed in Fig.~\ref{fig:masterplot}.

In the range $0^\circ\leq\Theta_\textrm{max}\leq 52^\circ$, solutions can be obtained both with the analytical ansatz and with the numerical optimization. We see that the maximum efficiency obtained with the ansatz stroke agrees with the optimal curve obtained in Sec.~\ref{sec:constr}, confirming the validity and form of the bounds on $\eta$ introduced by a restriction on $\Theta_\textrm{max}$ in our numerical approach. Beyond the value $\Theta_\textrm{max}=52^\circ$, the analytical ansatz provides an indication of the bound imposed on the efficiency by a weaker restriction on $\Theta_\textrm{max}$, in a domain not reachable by our numerical algorithm, and allows us to continuously link the results of the numerical optimization results to the theoretical optimal, but singular, stroke of $\eta=1/2$.  

Figure~\ref{fig:masterplot} presents thus the complete optimization diagram, where for each value of $\Theta_\textrm{max}$, the stroke has been optimized to lead to the largest possible swimming efficiency, and all values of $\eta$ from 0 to $1/2$ can  be obtained. {In particular, we observe that realistic values of the cilia-length-to-body-length ratio lead to the angular amplitude range $\Theta_\textrm{max}\approx 5$--$15^\circ$, which corresponds to optimal strokes of efficiency in the range $\eta\approx 3$--$6\%$ comparable to, if not slightly above, swimming efficiencies expected using flagellar propulsion \cite{brennen1977,lauga2009}.}

%%%%%%%%%%%%%%%%%%
% NEW SECTION: PROPERTIES
%%%%%%%%%%%%%%%%%%%

\section{Properties  of optimal strokes}
\label{sec:results}
In this section, we consider the optimal strokes obtained numerically using the unconstrained and constrained optimization algorithms of Secs.~\ref{sec:unconstr} and \ref{sec:constr}, and analyze their physical properties.

\subsection{Lagrangian vs. Eulerian quantities}
Two points of view can be adopted to analyze the surface deformation of an optimal squirmer. One can either consider the property of a {fixed point in the swimmer's co-moving frame} characterized by a polar angle $\theta$ (Eulerian formulation) or the property of a  {given material surface point} indexed by $\theta_0$ (Lagrangian formulation). A quantity $Q$ can then be understood and plotted either as a Lagrangian quantity $Q(\theta_0,t)$ or an Eulerian quantity $Q(\theta,t)$. To make this distinction, $\theta$ will be used in what follows for the fixed coordinate, and $\theta_0$ will be used as a Lagrangian label.

The mean position of a given material point $\theta_0=\mean{\theta}$ is one of the many possible Lagrangian labels. It is physically convenient and intuitive because it is expected to correspond roughly to the position of the cilia base. Note however that the configuration $\vartheta(\theta_0,t)=\theta_0$ for all $\theta_0$, corresponding to the case where all cilia are vertical at the same time, is never reached during the stroke period.

In our analysis, we will see that quite  different conclusions can be reached when considering  the same quantity from one point of view or  the other. This is illustrated below by looking at the stretching of surface elements. The Lagrangian formulation allows to address  the {individual} behavior of a given material point or cilia tip in time. In contrast, the Eulerian formulation is more adapted to study the global properties of the swimming stroke, determined by the collective behavior of the different surface points.

\subsection{Surface stretching and compression}
\begin{figure}
\begin{center}
\begin{tabular}{cc}
Optimal unconstrained  & Optimal constrained \\
$\eta\approx 22.2\%$, $\Theta_\textrm{max}\approx 52.6^\circ$, $\mean{U}=0.33$ & $\eta\approx 6.2\%$, $\Theta_\textrm{max}\approx10.6^\circ$ , $\mean{U}=0.058$ \\
\includegraphics[width=8cm]{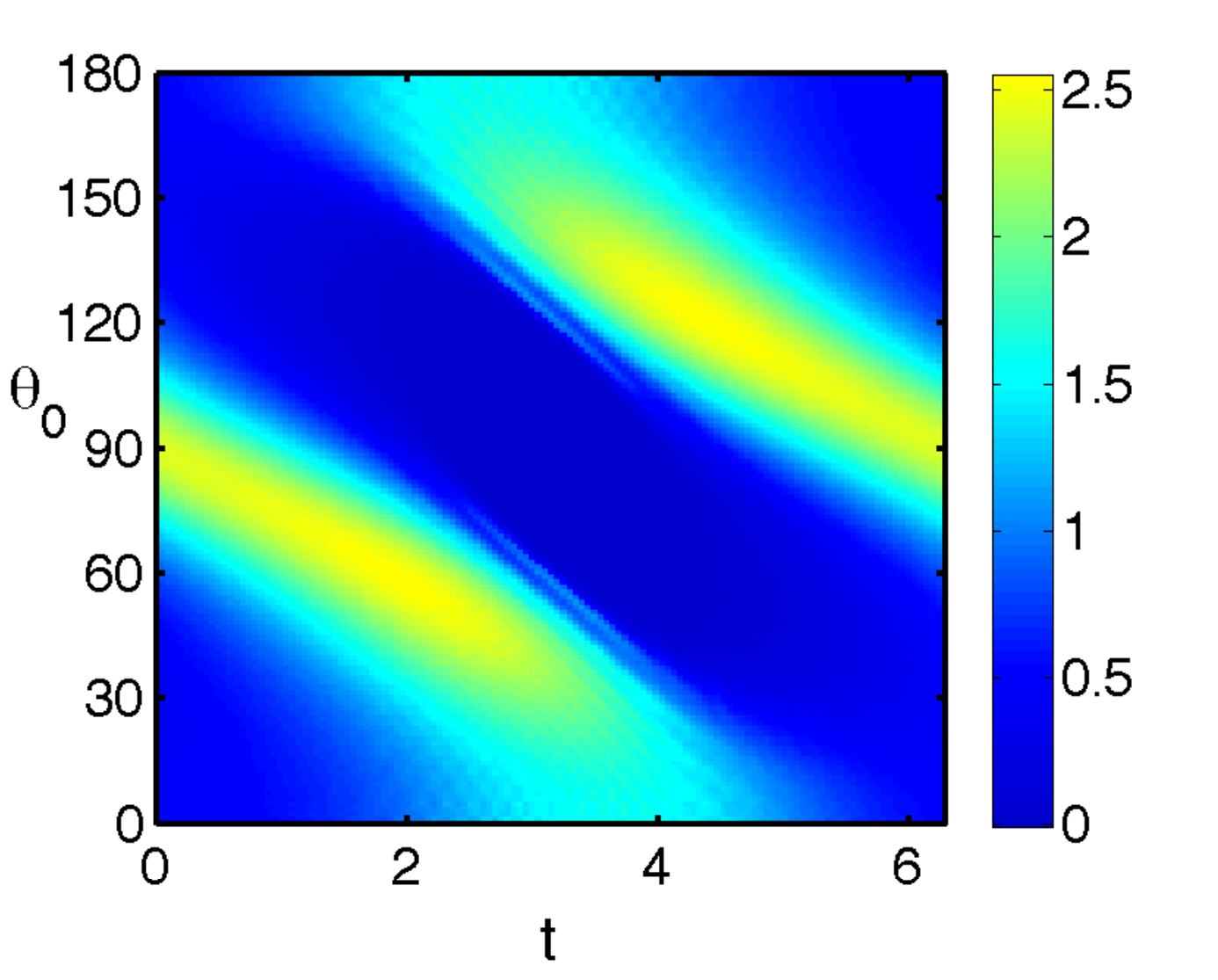} &
\includegraphics[width=8cm]{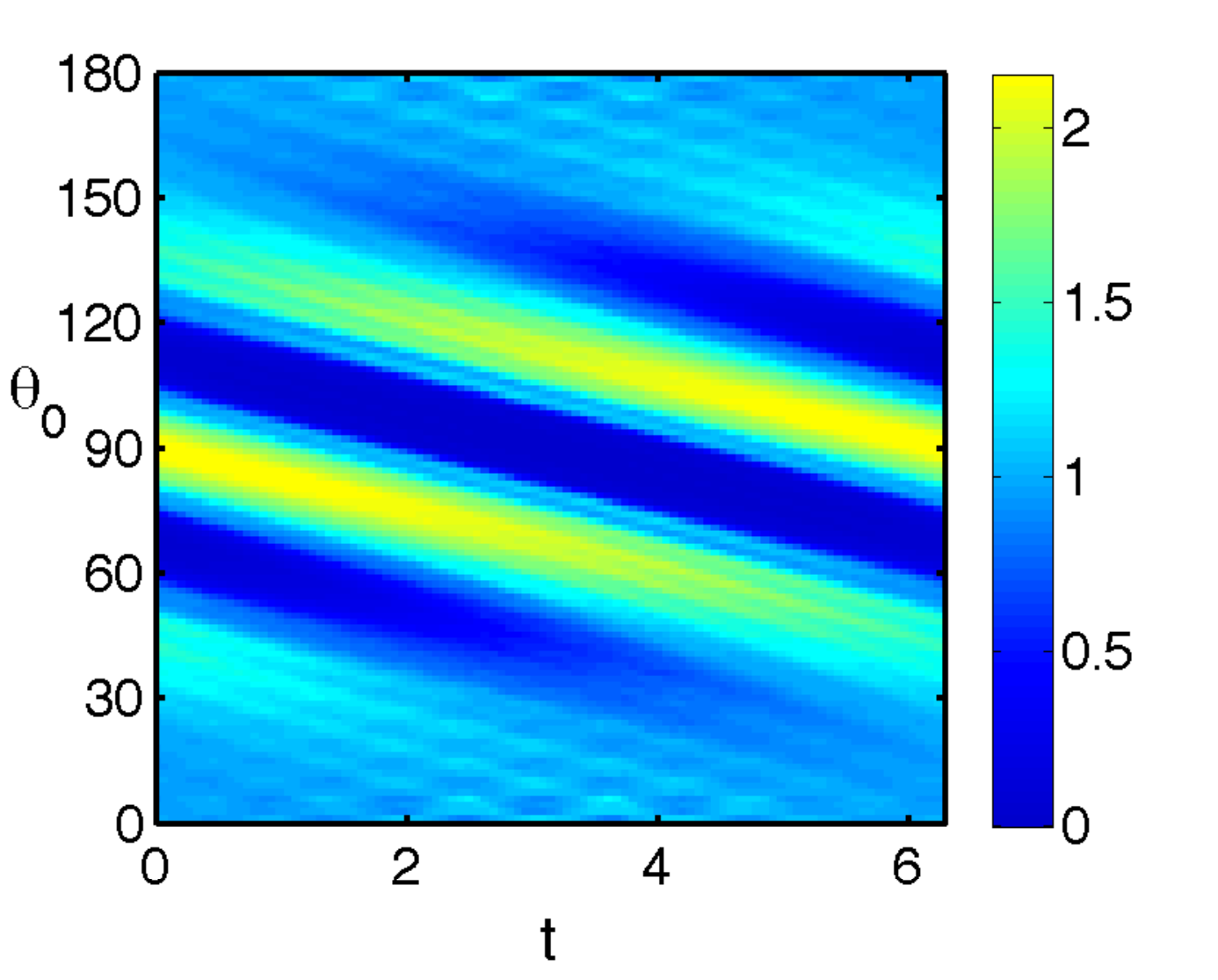} \\
\includegraphics[width=8cm]{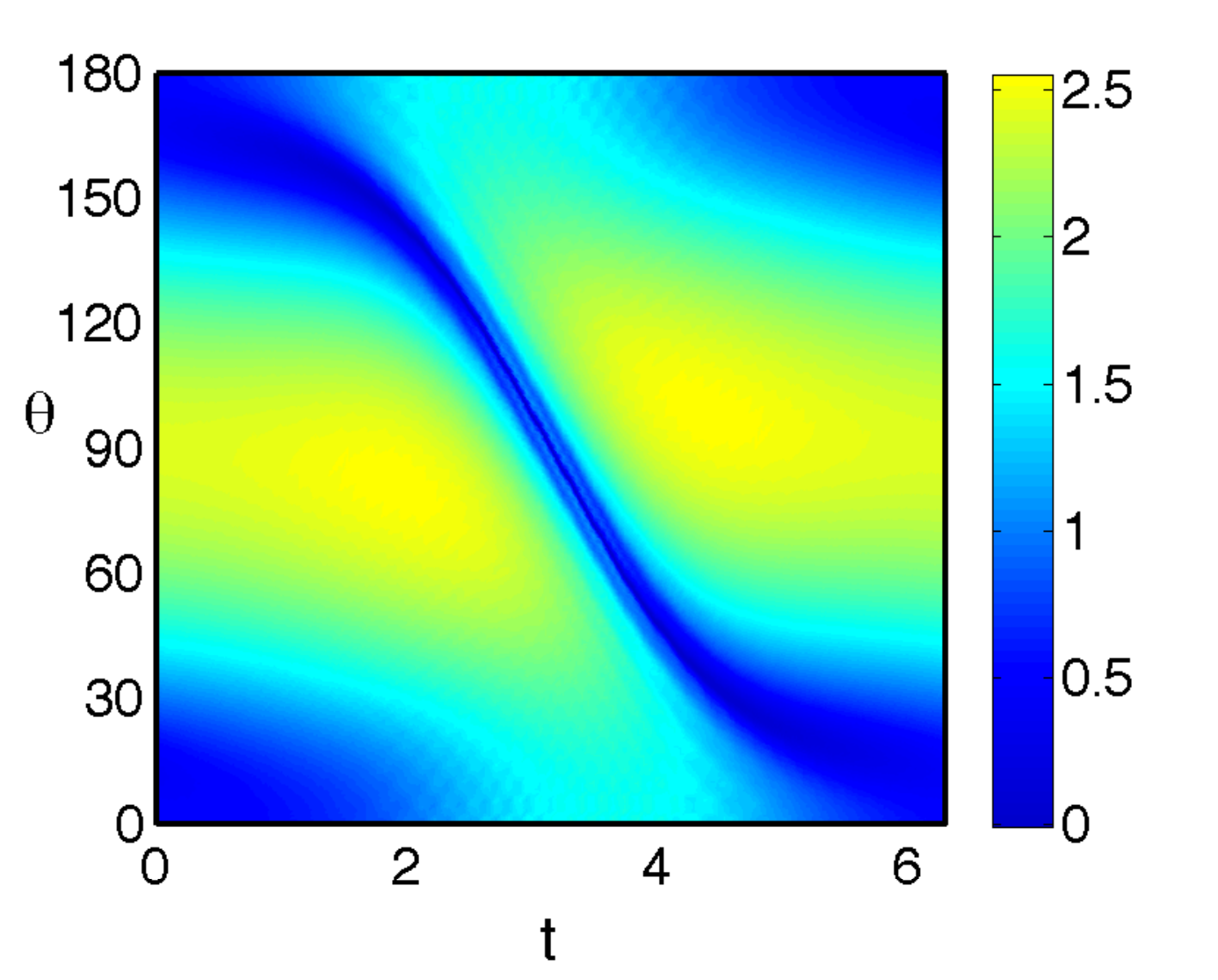} &
\includegraphics[width=8cm]{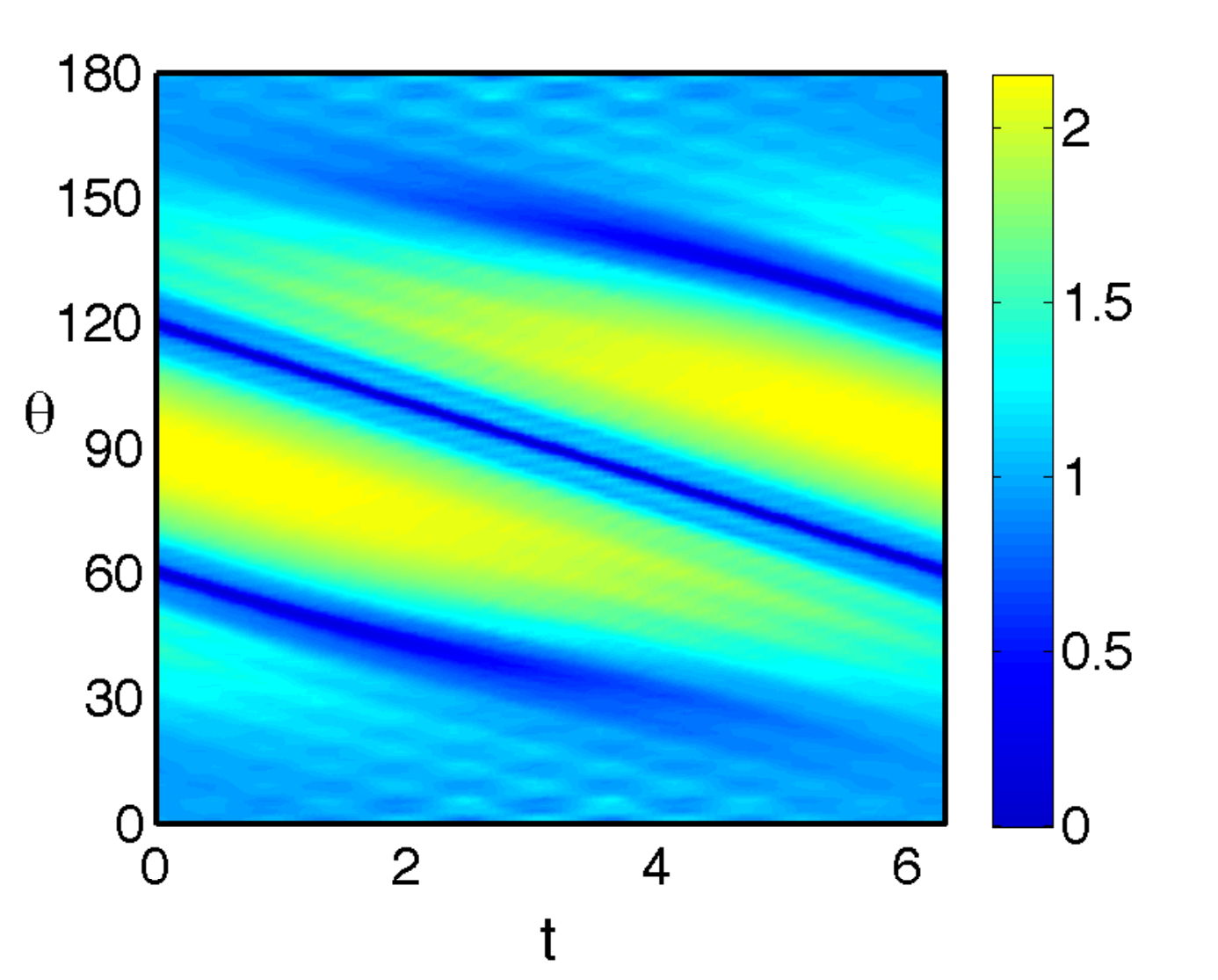} 
\end{tabular}
\caption{(color online) (Top) Lagrangian and (bottom) Eulerian map in the $(\theta,t)$ plane of the stretching parameter $
\mathcal{S}$ for the unconstrained swimming stroke (left) and the result of constrained optimization with efficiency $6.2\%$ and maximum displacement $10.6^\circ$ (right).}\label{fig:stretch}
\end{center}
\end{figure}

The difference in the displacement of two neighboring surface elements induces some periodic stretching and compression of the surface. This stretching can be quantified using $\mathcal{S}=\partial\vartheta/\partial\theta_0$, with $\theta_0$ the mean position of each surface point; {$\mathcal{S}>1$ (resp. $\mathcal{S}<1$) corresponds to a stretched (resp. compressed) surface in comparison with the reference (mean) configuration.} The stretching field $\mathcal{S}$ can be plotted either in Lagrangian coordinates as $\mathcal{S}(\theta_0,t)$ or in Eulerian coordinates as $\mathcal{S}(\theta,t)$, and both are shown in Fig.~\ref{fig:stretch} for two different strokes (the optimal unconstrained stroke with $\eta\approx22.2\%$ and the  constrained one with $\eta\approx6.2\%$).

One can see from the comparison of the top and bottom figures in Fig.~\ref{fig:stretch} that the Lagrangian or Eulerian point of view fundamentally changes the description of surface stretching. In the Lagrangian formulation, the stretching of a material particle appears symmetric, with a particle spending about as much time compressed than stretched. In the Eulerian approach, however, the conclusion is qualitatively different. In this case, a given location experiences a stretching of the surface for most of the  {period}, with compression being only achieved during the fast passage of the recovery shock. We therefore observe a strong symmetry-breaking at the organism level  in stretching and compression, whereas it is essentially symmetric at the individual cilium level. This symmetry-breaking, consistent with the discussion presented in Sec.~\ref{uresults}, 
 will be analyzed quantitatively in Sec.~\ref{sec:asymmetry}.

Note in addition that in Fig.~\ref{fig:stretch}, by comparing the stretching maps for the unconstrained and constrained strokes, we can observe additional changes   such as a smaller velocity of the shock in the constrained case. These properties will be discussed further in Sec.~\ref{sec:phase_vel}.

\subsection{Symmetry-breaking and collective behavior}
\label{sec:asymmetry}

\subsubsection{Individual (Lagrangian) motion asymmetry}
We first adopt the Lagrangian point of view to characterize the symmetry or asymmetry in the motion of individual surface elements. Quantitatively, different measures of this asymmetry can be chosen and we focus here on the following ones: \\
(1) {\emph{temporal asymmetry}, measured as the period fraction during which the material point is moving downward (resp. upward)
\begin{equation}
\label{eq:zeta1}
{\zeta_1(\theta_0)=\log\left(\frac{\tau\left(\displaystyle\pard{\theta}{t}(\theta_0,t)>0\right)}{\tau\left(\displaystyle\pard{\theta}{t}(\theta_0,t)<0\right)}\right);}
\end{equation}
}
(2)  {\emph{maximum velocity asymmetry}, defined as the ratio between the maximum velocities of the material point in each direction
\begin{equation}\label{eq:zeta2}
{\zeta_2(\theta_0)=\log\left|\frac{\textrm{Min}_{\,t}\left(\displaystyle\pard{\theta}{t}(\theta_0,t)\right)}{\textrm{Max}_{\,t}\left(\displaystyle\pard{ \theta}{t}(\theta_0,t)\right)}\right|;}
\end{equation}
}
(3) 
{\emph{individual trajectory eccentricity}, defined as the distance of the turning points to the mean position
\begin{equation}\label{eq:zeta3}
\zeta_3(\theta_0)=\log\left(\frac{\theta_\textrm{max}(\theta_0)-\theta_0}{\theta_0-\theta_\textrm{min}(\theta_0)}\right)\cdot
\end{equation}
}

\begin{figure}
\begin{center}
\begin{tabular}{cc}
\hspace{-.2cm}\subfigure[Lagrangian velocity map]{\includegraphics[height=6.5cm]{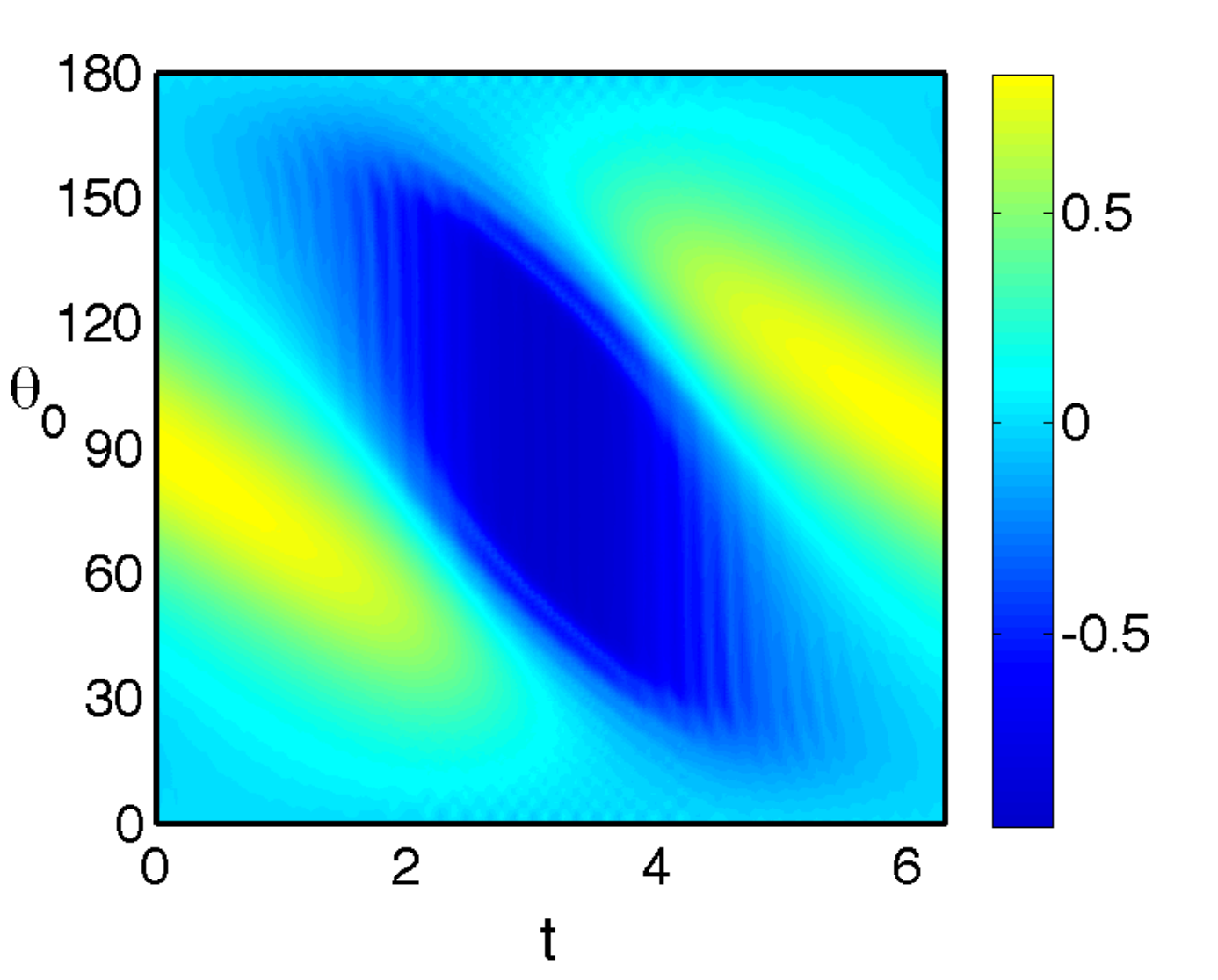} }&
 \hspace{-.6cm}\subfigure[Individual asymmetry]{\includegraphics[height=6.5cm]{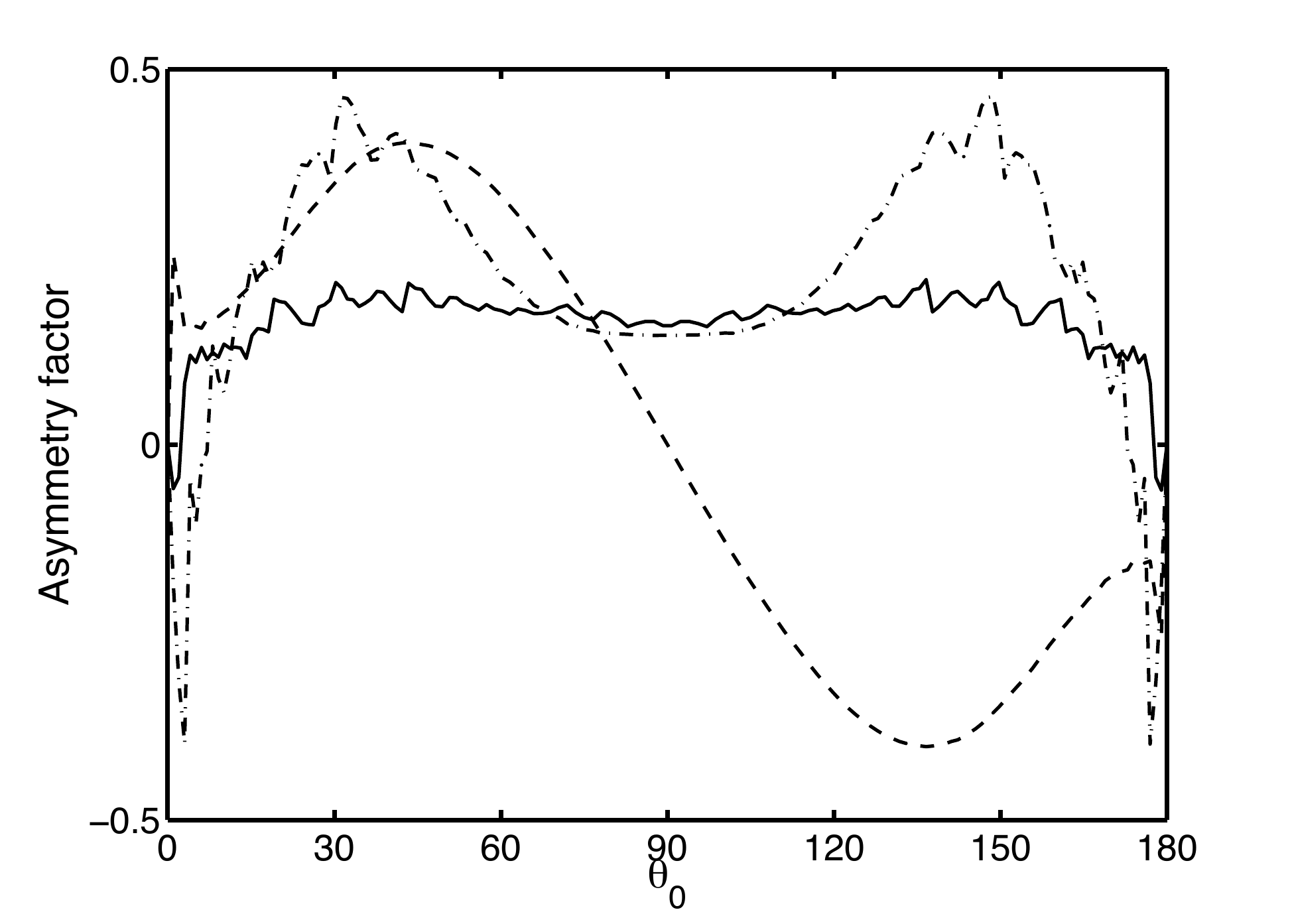}}
\end{tabular}
\caption{(color online) Asymmetry in the motion of individual surface points (Lagrangian approach).  (a) Lagrangian velocity map, $u_\theta(\theta_0,t)$. (b)  Asymmetry in the motion of individual material points measured as: the ratio of the period fraction with upward or downward velocity $\zeta_1$ (solid), the ratio of the maximum (in absolute value) positive and negative velocities $\zeta_2$ (dash-dotted) and the excentricity of individual trajectories $\zeta_3$ (dashed). All three measures are plotted for the unconstrained stroke ($\eta\approx22.2\%$ and $\Theta_\textrm{max}\approx 52.6^\circ$). }\label{fig:ind_asym}
\end{center}
\end{figure}

These three quantities are plotted as functions of the Lagrangian coordinate,  $\theta_0$, on Fig.~\ref{fig:ind_asym}. A logarithmic scaling is used so that $\zeta_p=0$ corresponds to the symmetric situation and opposite sign $\zeta_p$ corresponds to the same amplitude of asymmetry in opposite directions. Regardless of the chosen asymmetry measure, we see that it vanishes near the poles, and  the equatorial points correspond to local minima of the asymmetry. The maximum individual asymmetry is reached around $\theta_0\approx40$--$45^\circ$, and this maximum is more pronounced for $\zeta_2$ and $\zeta_3$ measuring a displacement asymmetry than for the temporal asymmetry factor, $\zeta_1$. We also see that $\zeta_1$ and $\zeta_2$ are positive everywhere (except near the poles, where the very small displacements make the asymmetry measurements irrelevant), while $\zeta_3$ has opposite sign depending on the hemisphere, showing an eccentricity of the individual trajectories directed toward the equator.

The value of the temporal asymmetry, $\zeta_1$, is rather small and uniform for individual particles ($\zeta_1\approx0.2$ corresponds to a $20\%$ asymmetry in the velocity sign) confirming the near-symmetry observed in the map of surface velocity displayed in Fig.~\ref{fig:ind_asym}a. Individual cilia  spend thus about the same time moving in either direction, but a slight asymmetry is observed, with about  $20\%$ more time spent performing the effective stroke (positive $\partial\theta/\partial t$). As a consequence, the maximum velocity achieved is greater during the recovery stroke (positive $\zeta_2$).

\subsubsection{Group (Eulerian) asymmetry}

\begin{figure}
\begin{center}
\begin{tabular}{cc}
\subfigure[Eulerian velocity map]{\includegraphics[height=6.5cm]{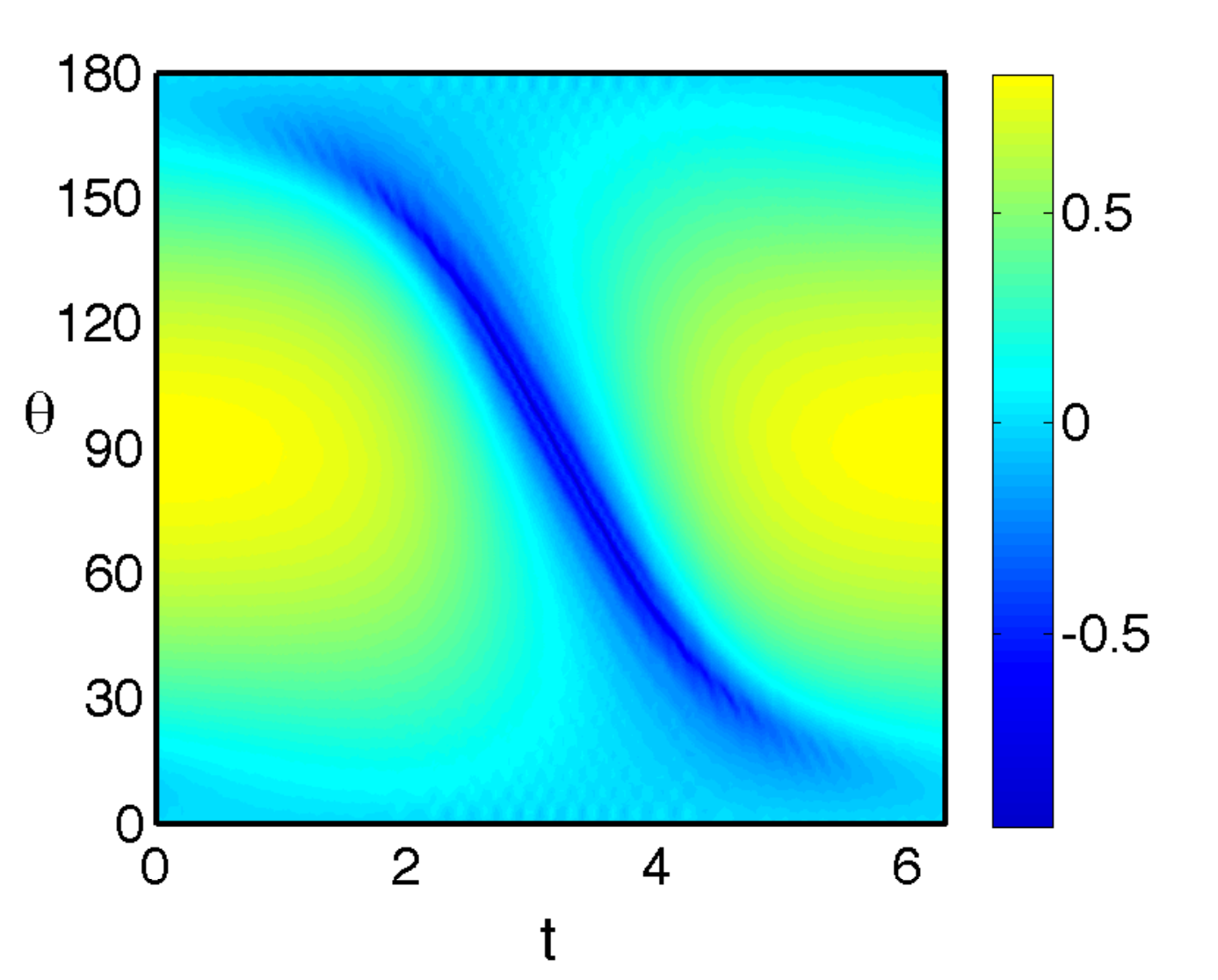}} &
\subfigure[Eulerian temporal asymmetry ratio]{\includegraphics[height=6cm]{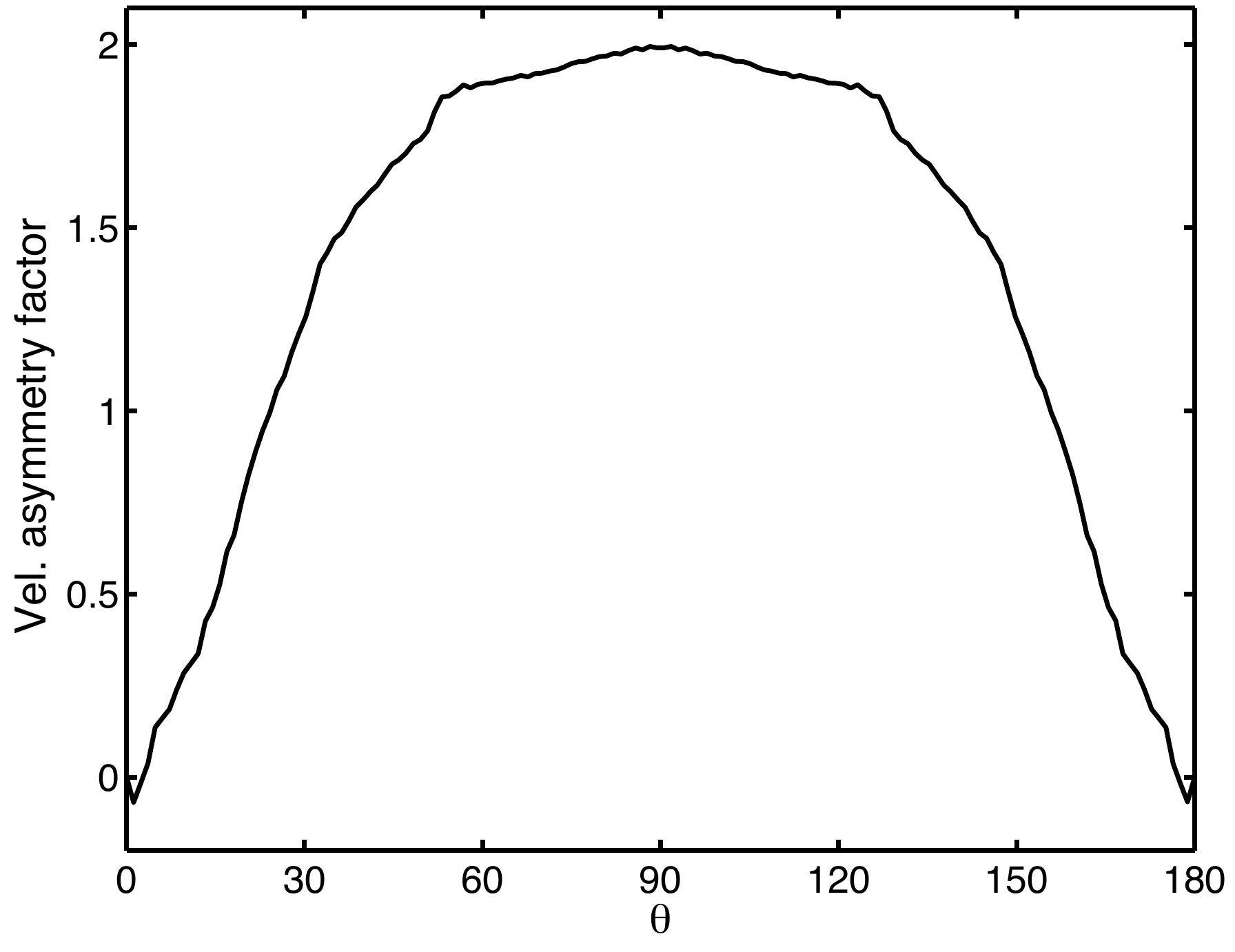}}
\end{tabular}
\caption{(color online) (a) Surface velocity map of the Eulerian velocity, $u_\theta(\theta,t)$. (b) Eulerian asymmetry measured as the ratio $\tilde\zeta_1(\theta)$ (Eq.~\ref{eq:zeta1b}) between the period fractions where the velocity at a fixed point is directed upward vs. downward. Data plotted for the optimal unconstrained stroke  ($\eta\approx22.2\%$ and $\Theta_\textrm{max}\approx52.6^\circ$). }\label{fig:grp_asym}
\end{center}
\end{figure}

Although individual material points display a near-symmetric motion over the whole period, the collective behavior of the different surface elements and the phase difference in their motion leads to a symmetry breaking at the organism level. This was illustrated above by considering the stretching/compression parameters $\mathcal{S}$ in both Lagrangian and Eulerian coordinates, and observing that the near-symmetric Lagrangian patterns  show significant  Eulerian asymmetry.

A similar disparity is apparent between the Lagrangian (Figs.~\ref{fig:ind_asym}a) and Eulerian (Figs.~\ref{fig:grp_asym}b) tangential velocity maps. The Eulerian asymmetry can be quantified using the normalized temporal asymmetry ratio  $\tilde\zeta_1$, the Eulerian equivalent to $\zeta_1$  (Eq.~\ref{eq:zeta1}):
\begin{equation}\label{eq:zeta1b}
\tilde\zeta_1(\theta)=\log\left(\frac{\tau\left(u(\theta,t)>0\right)}{\tau\left(u(\theta,t)<0\right)}\right),
\end{equation}
and the results are shown on Fig.~\ref{fig:grp_asym}b. In comparison to Fig.~\ref{fig:ind_asym}b, we observe a significant increase of this asymmetry factor in comparison to the individual asymmetry measurement. While individual material points spend only a maximum of $20\%$ more time performing the effective vs. the recovery stroke, a fixed Eulerian location on the swimmer near the equator is part of the effective stroke for $90\%$ of the period. Note that while the results of Figs.~\ref{fig:ind_asym} and \ref{fig:grp_asym} are shown for the unconstrained optimal stroke, similar behavior is observed for the strokes with constrained displacement amplitudes. 

In optimal strokes, although the motion of individual  points on the swimmer surface is roughly symmetric, the collective behavior of all surface points display a global symmetry-breaking between the two phases of the swimming stroke (effective and recovery). As discussed in Sec.~\ref{sec:ansatz},  such breaking of the front-back symmetry  in the squirming dynamics is essential to efficient swimming in order to  ensure that the effective stroke dominated by the swimming mode covers most of the $(\theta,t)$-plane.

\subsection{Modal decomposition of the optimal stroke and wave propagation}
\label{sec:phase_vel}
The emergence of global asymmetry from near-symmetric individual motion typically occurs when the individual displacement is harmonic with a spatially-varying phase. For example, trajectories of the form 
\begin{equation}\label{eq:waveform}
\theta=\vartheta(\theta_0,t)=\theta_0+a(\theta_0)\cos(t-\theta_0/v_\theta)
\end{equation}
would correspond to exactly symmetric individual trajectories, but for sufficiently large amplitude $a(\theta_0)$, non-symmetric collective behavior is observed,  thereby breaking the symmetry between upward and downward strokes (see Fig.~\ref{fig:shift}). 

\subsubsection{Complex Empirical Orthogonal Function (CEOF) decomposition}
The symmetry-breaking identified in the optimal strokes is associated with a propagating wave reminiscent of the metachronal waves observed on the surface of many ciliated organisms \cite{brennen1977}. Here, we confirm the existence of this wave and quantify its properties through the use of a complex modal decomposition \cite{michelin2008}. This method is based on the extension of classical Empirical Orthogonal Functions (EOF -- also known as Proper Orthogonal Decomposition) to complex variables, and is well-designed to study propagating two-dimensional wave patterns \cite{barnett1983}. 

In the classical EOF method, a function of space and time, {$f(\theta_0,t)$}, is decomposed in a series of modes
\begin{equation}
{f(\theta_0,t)=\displaystyle\sum_n\tilde{a}_n(\theta_0)\tilde{b}_n(t),}
\end{equation}
designed in such a way that truncation of the infinite sum at any order $N$ minimizes the {$L_2$-norm of the remainder.} The Complex EOF decomposition performs the classical EOF decomposition on the Hilbert transform  {$F(\theta_0,t)$} of  {$f(\theta_0,t)$}. This technique is used here to study the displacement of surface points from their mean position $\theta_0$, which is decomposed as
\begin{equation}
\label{eq:CEOF}
\vartheta(\theta_0,t)-\theta_0=\Real\left[\sum_nA_n(\theta_0)\overline{B_n(t)}\right]=\sum_n a_n(\theta_0)b_n(t)\cos(\phi_n(\theta_0)-\psi_n(t)).
\end{equation}
\begin{figure}
\begin{center}
\includegraphics[width=15cm]{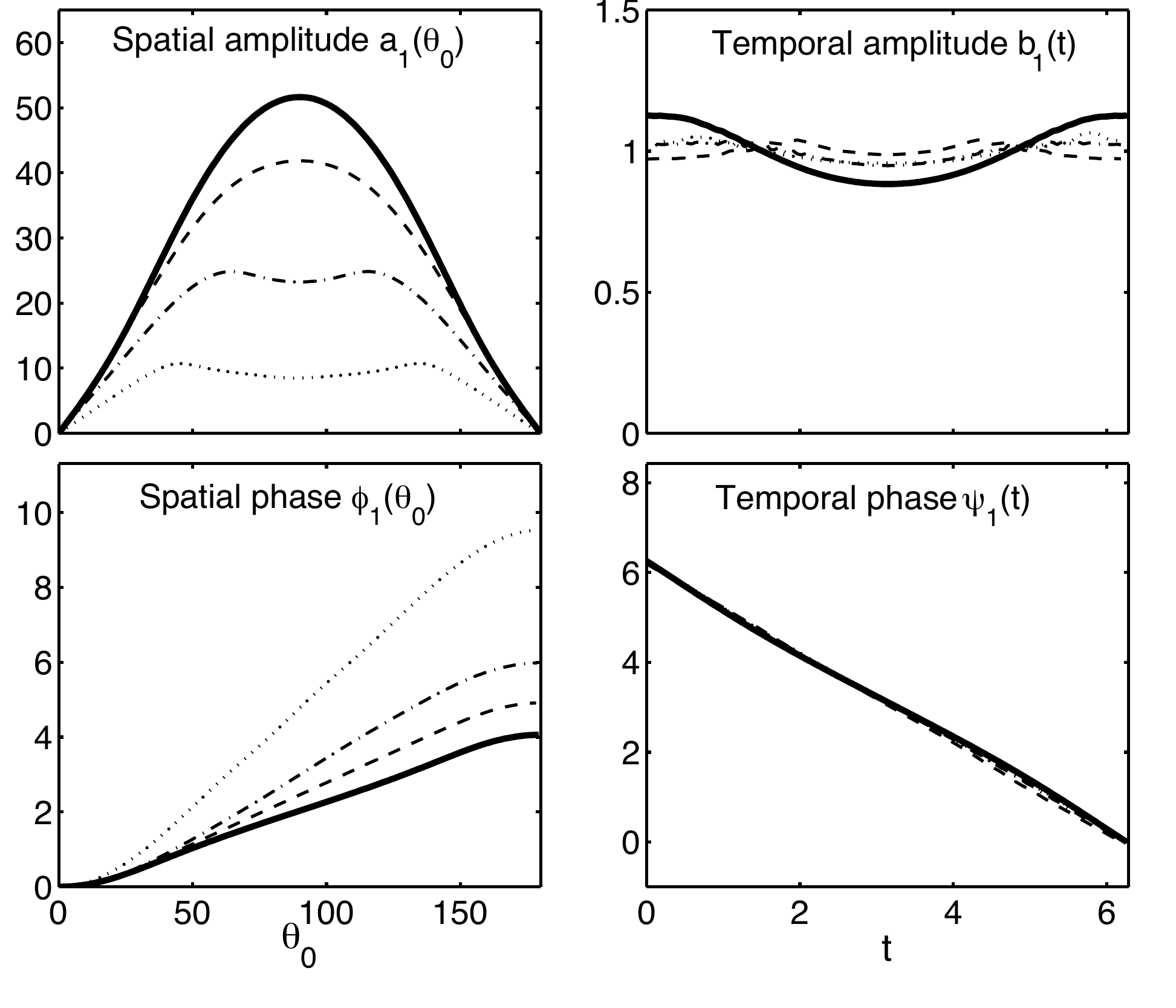}
\caption{Amplitude and phase characteristics of the dominant Complex Empirical Orthogonal Function (CEOF) mode
  in the unconstrained swimming stroke of maximum
  efficiency (thick solid line -- $\Theta_\textrm{max}\approx52.6^\circ$ and $\eta\approx22.2\%$) and in constrained optimal strokes (thin lines):  $\Theta_\textrm{max}\approx44^\circ$ and $\eta\approx21\%$ (dashed),  $\Theta_\textrm{max}\approx25^\circ$ and $\eta\approx12.5\%$ (dash-dotted), $\Theta_\textrm{max}\approx11^\circ$ and $\eta\approx6.8\%$  (dotted).}\label{fig:CEOF_ev}
\end{center}
\end{figure}

This method is particularly powerful for our time-periodic problem. For all the optimal swimming strokes obtained in Secs.~\ref{sec:unconstr} and \ref{sec:constr}, the first mode is so dominant that the error introduced by restricting the sum in Eq.~\eqref{eq:CEOF} to the first mode is found to be less than $1$--$2\%$. The characteristics (temporal and spatial amplitude, and phase functions) of this first mode are shown in Fig.~\ref{fig:CEOF_ev} for both the unconstrained and constrained optimal strokes with decreasing displacement amplitude. Note that the definition of $a_n$ and $b_n$ in Eq.~\eqref{eq:CEOF} is not unique, and one is free to choose a normalization for one of these functions. We chose to normalize the temporal amplitude so that $\mean{b_1}=1$. Similarly, the phase functions $\phi_1$ and $\psi_1$ are defined up to an arbitrary additive constant.

\subsubsection{Propagating wave mode and phase velocity}

In all cases plotted on Fig.~\ref{fig:CEOF_ev}, the temporal phase $\psi_1(t)$ is found to be a linear function of time with unit slope, which is consistent with the $2\pi$-stroke period. The spatial phase $\phi_1(\theta_0)$ varies also linearly, except in narrow regions near the poles. Outside these polar regions, the first CEOF mode is therefore a progressive wave of phase velocity 
\begin{equation}
V_\theta=\dot{\psi}_1/\phi_1',
\end{equation}
where dot and prime denote respectively a differentiation with respect to $t$ and $\theta_0$. This phase velocity   can be understood either as an angular velocity or a linear tangential velocity along the swimmer surface (in radii per unit time). Equivalently, the wavelength $\lambda$ of the optimal surface deformation is defined as 
\begin{equation}
\lambda=2\pi/\phi_1'=2\pi\,|V_\theta|,
\end{equation}
as the wave period is equal to $2\pi$. 
\begin{figure}
\begin{center}
\includegraphics[width=13cm]{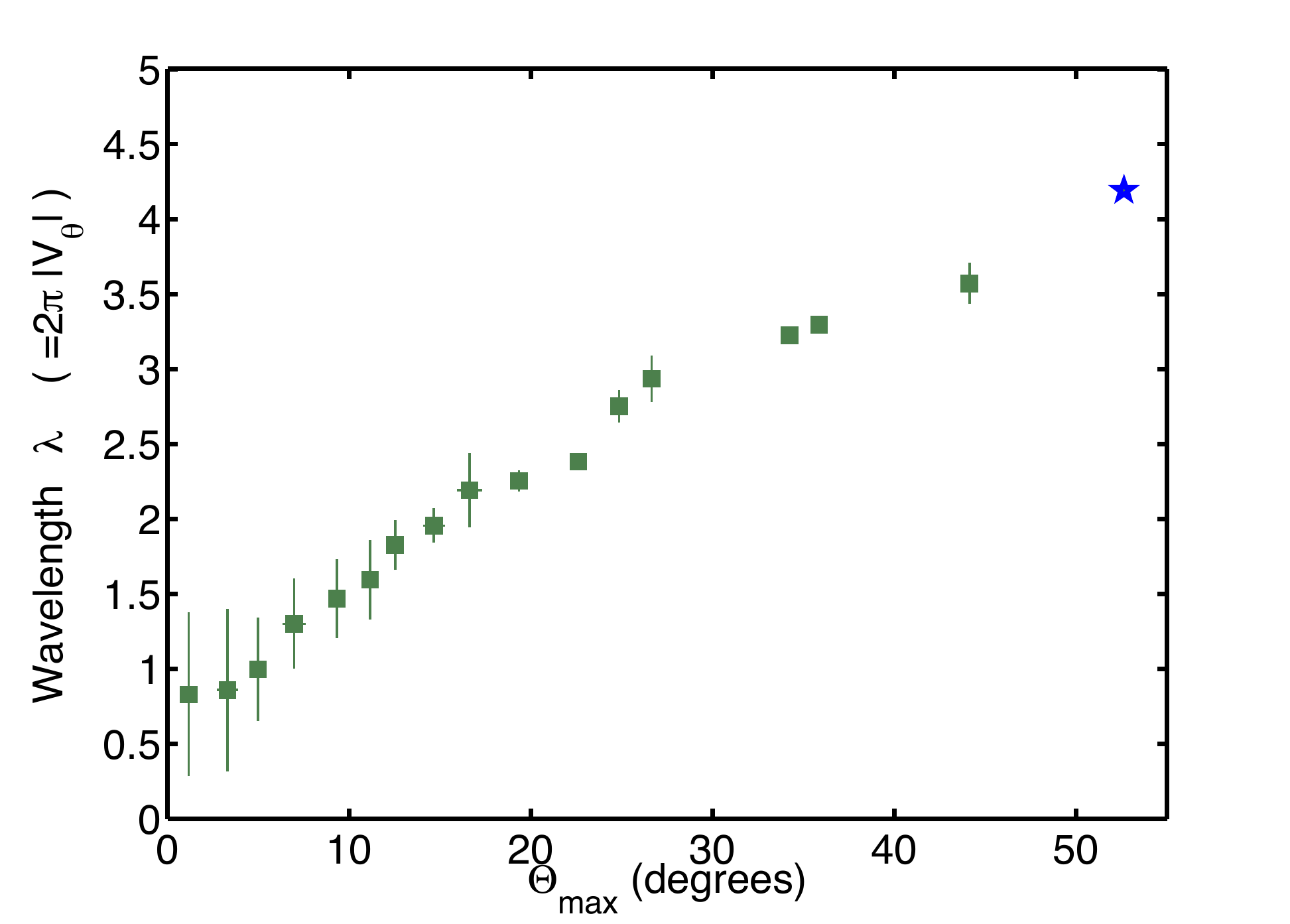}
\caption{(color online) {Variation of the wavelength $\lambda$ (or phase velocity $|V_\theta|=\lambda/2\pi$) measured along the surface of the spherical swimmer, with} the maximum angular displacement,  $\Theta_\textrm{max}$.  {The blue star corresponds to the unconstrained optimal swimming stroke of Sec. \ref{sec:unconstr}.}}\label{fig:phase_vel}
\end{center}
\end{figure}

{As the constraint on the maximum displacement $\Theta_\textrm{max}$ is stiffened (and the swimming efficiency is reduced), the magnitude of the wave velocity $V_\theta$ and wavelength $\lambda$ are reduced ($\phi_1'$ increases). The modal decomposition is used to determine the phase velocity for all the optimal strokes presented in Sec.~\ref{sec:constr}, and the dependance of $\lambda$ (or $V_\theta$) with $\Theta_\textrm{max}$ is shown on Fig.~\ref{fig:phase_vel}. 
The maximum wave velocity is reached for the stroke of maximum efficiency (and displacement), for which the phase velocity magnitude is approximately equal to 
$0.67$ (in radians) corresponding to a wavelength of about four radii (or two body lengths). For the stroke with smallest amplitude shown in Fig.~\ref{fig:CEOF_ev}, the phase velocity magnitude is reduced to about $0.27$ corresponding to a wavelength of about $1.5$ radii. Note that this reduction of the phase velocity and wavelengths with the stiffness of the constraint on the maximum displacement was also apparent  by comparison of the  slope of the recovery shock  in Figs.~\ref{fig:theta_plot_reference} and \ref{fig:theta_plot_constrained}.}

The phase velocity of the wave is negative in $\theta$ (or positive in $x$) corresponding to a wave always traveling toward the north pole, in the same direction as the swimming velocity, and thus in the opposite direction to the effective stroke. Such metachronal waves are known as antiplectic \cite{blake1974b,brennen1977}. This result is consistent with the Taylor's swimming sheet model  for which the swimming velocity and the wave velocity have same direction for  tangential deformations  \cite{blake1971b,childress1981,stone1996}, whereas  normal displacements lead to wave and swimming velocities of opposite directions (symplectic metachronal waves) \cite{taylor1951,childress1981}.

Note finally that the waves quantified above are defined in terms of $\theta_0$ and $t$. The phase velocity is thus measured using the Lagrangian label rather than the actual instantaneous position of the cilia. Because the amplitude of beating is not small, this introduces a small difference between the quantitative measure of the phase velocity we propose and the phase velocity visually assessed, for example, from Fig.~\ref{fig:theta_plot_reference}. A formal wave analysis in Eulerian coordinates leads to insignificant changes to the conclusions of the Lagrangian analysis.

\subsubsection{Mode amplitude}

The structure of the dominant mode presented on Fig.~\ref{fig:CEOF_ev} also characterizes the amplitude  of the tangential motion    through the spatial and temporal amplitude functions $a_1(\theta_0)$ and $b_1(t)$. The temporal amplitude was normalized to unit mean value so that the spatial amplitude provides the physical value of the amplitude  of displacement. We see that the departure from unity of $b_1$ is small confirming the structure of the dominant mode as a traveling wave with spatial variations of its amplitude.

The spatial amplitude $a_1$ is observed to match very well the function $\Theta(\theta_0)$ obtained for each value of $\theta_0$ as the amplitude of motion of the particular Lagrangian point. For the unconstrained stroke it is maximal at the equator. As the constraint on maximum displacement is stiffened, the spatial amplitude is reduced ($\Theta_\textrm{max}$ in Figs.~\ref{fig:eff_th} and \ref{fig:masterplot} corresponds to the peak of this curve) and its profile is flattened. As strokes with smaller $\Theta_\textrm{max}$ are considered, the displacement amplitude of individual surface elements display  smaller disparities  between equatorial points and the rest of the surface. Due to the spherical symmetry, the shape function $a_1$ must go to zero near the poles and can therefore not be constant over the whole surface of the swimmer. However, we notice that for the most constrained case plotted in Fig.~\ref{fig:CEOF_ev} the spatial amplitude is close to $\Theta_\textrm{max}$ for $30^\circ\leq\theta_0\leq150^\circ$. This result is particularly interesting as the strokes of  ciliated microorganisms  correspond to the lower end of the $\Theta_\textrm{max}$-range considered here, where the shape function is the flattest. In such cases, the surface elements display, apart from the poles,  a constant amplitude along the swimmer. This is consistent with the physical intuition that individual cilia do not ``know" whether they are located near the pole or the equator and therefore, the beating properties (other than the phase) should be expected to be constant over the surface of the organism.

\subsubsection{Scaling of the swimming-to-wave velocity ratio} 
\begin{figure}
\begin{center}
\includegraphics[width=12cm]{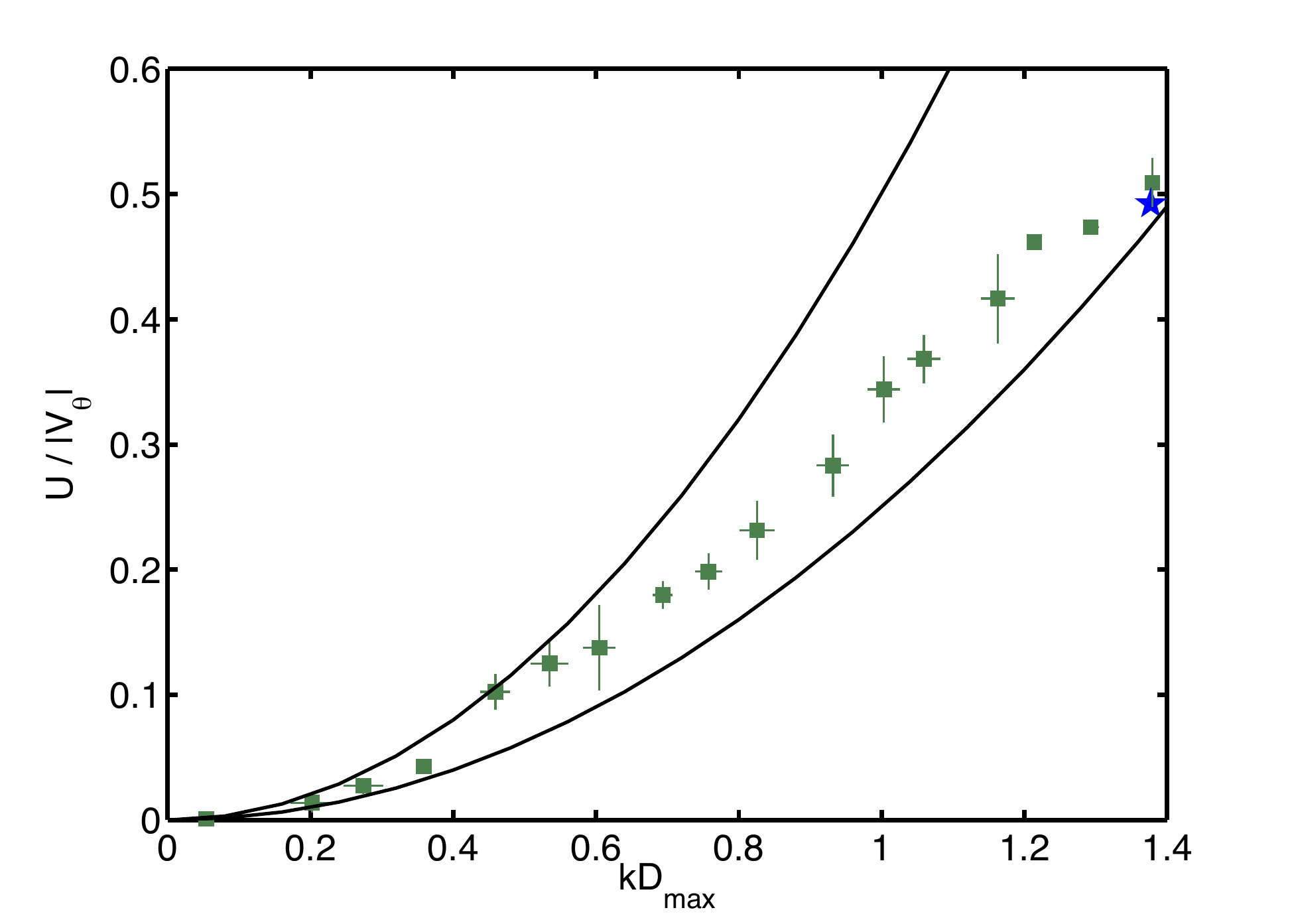}
\caption{(color online) Ratio of the swimming velocity to the wave speed, $U/V_\theta$, as a function of the maximum tangential displacement multiplied by the optimal wave number. As in other figures, results are binned and error bars are given both horizontally and vertically.  {The blue star corresponds to the unconstrained optimal swimming stroke of Sec. \ref{sec:unconstr}.} The two solid lines correspond respectively to $U/V_\theta=(kD_\textrm{max})^2/4$ and $U/V_\theta=(kD_\textrm{max})^2/2$ \cite{brennen1977}.}\label{fig:Uc_vs_kl}
\end{center}
\end{figure}

In the limit of small amplitude displacements, the swimming sheet model first introduced by Taylor shows that the swimming velocity (scaled by the wave propagation speed) is a quadratic function of the displacement amplitude \cite{taylor1951,childress1981}. This scaling also applies to surface motions of larger amplitudes, and for a large variety of ciliated microorganisms, it has been established that $U/c\sim\alpha (kl)^2$ with $\alpha\sim 0.25$ -- $0.5$ and  {$c$ is the wave velocity, $l$ the cilia length and $k$ the wave number} (see Fig.~20 in Ref.~\cite{brennen1977}).  We plot in Fig.~\ref{fig:Uc_vs_kl} the variations of $U/|V_\theta|$ with $kD_\textrm{max}$ where  {$D_\textrm{max}=\Theta_\textrm{max}$ is the maximum linear displacement amplitude and $k=1/|V_\theta|$ is the wavenumber of the optimal strokes  (the stroke period and swimmer's radius have been normalized to $2\pi$ and $1$ respectively)}. We observe the same quadratic scaling of $U$ with $kD_\textrm{max}$, with a numerical prefactor which lies in the same range.

A few comments must however be made. First, the results of  Fig.~20 in Ref.~\cite{brennen1977} use the length $l$ of an individual cilium rather than the maximum displacement. For most ciliated organisms, a good estimate of the cilia length is obtained as $l\sim 2D_\textrm{max}$ \cite{brennen1977}. In addition, most ciliated organisms reviewed in Ref.~\cite{brennen1977} are not spherical but actually elongated in the swimming direction (such as the one shown in Fig.~\ref{picture}). For such an organism, the wave velocity along the surface is very close to its component along the swimming direction, while these quantities differ by a factor $\pi/2$ on a sphere. The factor $U/V_x$ is therefore increased if the projection $V_x$ of $|V_\theta|$ along $\eb_x$ is considered. As a result of these rescalings, one would find that $U/c\approx0.1$ -- $0.2\,k^2l^2$, which is  close but not exactly similar to the  $U/c\approx 0.25$ -- $0.5\,k^2l^2$ obtained in Ref.~\cite{brennen1977}. 
Additionally, we observe that the optimal strokes correspond to waves traveling faster than the swimming velocity, while many microorganisms show much slower wave speeds (resulting in $U/|V_\theta|>1$). This limitation was already mentioned in Ref.~\cite{blake1974b} and points to 
the limits of the envelope model.

%%%%%%%%%%%%%%%%%%%%%%
% CONCLUSIONS
%%%%%%%%%%%%%%%%%%%%%%

\section{Discussion}
\label{sec:conclusions}

In this paper we have considered a spherical envelope model (so-called squirmer) to investigate energetics in cilia dynamics and locomotion. Allowing only tangential but time-periodic deformations, we have used an optimization method based on a variational approach to  derive computationally the stroke leading to the largest swimming efficiency. The optimal stroke was shown to display weak Lagrangian asymmetry, but strong Eulerian asymmetry, indicative of symmetry-breaking at the whole-organism level, but not at the level of individual cilia. 
We then added a constraint in the optimization approach to penalize large-amplitude deformation of the surface, together with a numerical ansatz, and derived a complete optimization diagram where all values of the swimming efficiency between 0 and 50\% (mathematical upper bound) could be reached. We were also able to construct a swimmer which is 50\% efficient, although it is mathematically singular. The deformation kinematics of the optimal strokes were always found to be wave-like, which we analyzed using a formal modal decomposition approach, and is reminiscent of metachronal waves in cilia arrays.

Our optimization procedure always leads to swimming strokes displaying antiplectic waves. 
This is, in fact, an intrinsic feature (and admittedly, a limitation) of the squirmer envelope model. In a real ciliated organism, the effective and recovery strokes correspond to very different shapes of the individual cilium  \cite{blake1974b,brennen1977}. This asymmetry defines the effective and recovery strokes, independently from the coordination of neighboring cilia. In the present envelope model, such an asymmetry is not represented, as only the  tangential displacement of cilia tips is prescribed, and thus  the effective or recovery nature of the stroke is determined by the collective behavior. 

Previous calculations in the limit of small deformations and harmonic waves for the swimming sheet model \cite{blake1971b,brennen1977} or the squirmer model \cite{stone1996} have shown that the swimming velocity is always oriented in the same direction as the propagating wave. A physical argument leading to the same conclusion in the squirmer geometry can be proposed as follows. In the envelope model with tangential displacements, the half-stroke in the direction of the wave propagation corresponds to a compression of the surface. The definition of the average swimming velocity from Eq.~\eqref{eq:alphn_comp} as a geometrically weighted-average of the Eulerian surface velocity identifies the effective stroke (of opposite direction to the swimming velocity) as the stroke where the surface is stretched because it occupies a greater part of the velocity map in the Eulerian $(\theta,t)$-plane (see Fig.~\ref{fig:grp_asym}a). The wave and the effective stroke therefore have opposite directions, and the metachronal wave is antiplectic. 

This can also be seen mathematically. For the general swimming strokes in Eq.~\eqref{eq:waveform}, we can obtain using Eq.~\eqref{eq:alphn_comp} that
\begin{equation}\label{eq:wavedir}
\mean{U}\times v_\theta=-\frac{1}{4\pi}\int_0^\pi\int_0^{2\pi}\Big\{a(\theta_0)g'(t-\theta_0/v_\theta)\sin\big[\theta_0+a(\theta_0)g(t-\theta_0/v_\theta)\big]\Big\}^2\dd t\,\dd\theta_0<0.
\end{equation}
As a result, a deformation wave traveling from the south to the north pole ($v_\theta<0$) imposes a northward swimming velocity {in the same direction as the wave}. Although we do not prove this result here for arbitrary periodic functions $\vartheta(\theta_0,t)$, it appears to be a general result. The simplified assumptions of the envelope model used here therefore prevent us to conclude on the relative efficiency of both types of metachronal waves. 

Our work was but a first attempt at an optimization approach to cilia dynamics, and could be extended in a variety of ways. First, and foremost, one should allow deformations normal to the swimmer surface to take place. In that case the swimmer would display time-periodic shape changes, and the  spherical-harmonics framework will no longer be applicable. A full numerical approach would thus have to be implemented to derive the instantaneous swimming speed and energetics. Another possible extension would allow for non-axisymmetric deformation to occur, in which case the swimming kinematics would also be including a rotation. The dissipation arguments presented in Ref.~\cite{stone1996} suggest that any such rotation would be sub-optimal, but it would be instructive to obtain that result from an optimization approach (see also \cite{shapere1987,shapere1989}). The straightforward extension of our results to other shapes with tangential deformations, in particular prolate spheroid-like, would also be biologically relevant. Finally, one should consider other biologically-relevant transport quantities, such as the flux of nutrients transported by the swimming-induced flow. 

\section*{Acknowledgments}
Funding  by the National Science Foundation (grant  CBET-0746285 to E.L.) is gratefully acknowledged .

\bibliographystyle{unsrt}
\bibliography{cilia-refs.bib}
\end{document}